%% file: main.tex
\documentclass[sigconf,urlbreakonhyphens=false,screen]{acmart}

\usepackage{acmart-taps} 
\usepackage{booktabs} 
\usepackage[ruled]{algorithm2e}
\usepackage{fancyhdr}
\usepackage{soul}
\usepackage{color}
\usepackage{subfig}
\usepackage{bbding}
\usepackage{colortbl}

\SetAlFnt{\small}
\SetAlCapFnt{\small}
\SetAlCapNameFnt{\small}
\SetAlCapHSkip{0pt}

\AtBeginDocument{%
  \providecommand\BibTeX{{%
    \normalfont B\kern-0.5em{\scshape i\kern-0.25em b}\kern-0.8em\TeX}}}

\setcopyright{acmcopyright}

\copyrightyear{2023} 
\acmYear{2023} 
\setcopyright{acmlicensed}
\acmConference[CHI '23]{Proceedings of the 2023 CHI Conference on Human Factors in Computing Systems}{April 23--28, 2023}{Hamburg, Germany}
\acmBooktitle{Proceedings of the 2023 CHI Conference on Human Factors in Computing Systems (CHI '23), April 23--28, 2023, Hamburg, Germany}
\acmPrice{15.00}
\acmDOI{10.1145/3544548.3581024}
\acmISBN{978-1-4503-9421-5/23/04}

\begin{document}
\title{AnisoTag:~3D~Printed~Tag~on~2D~Surface~via~Reflection~Anisotropy}

\author{Zehua Ma}
\orcid{0000-0002-8153-341X}
\email{mzh045@mail.ustc.edu.cn}
\affiliation{%
  \institution{University of Science and Technology of China}
  \city{Hefei}
  \country{China}
}

\author{Hang Zhou}
\authornote{The corresponding author.}
\orcid{0000-0001-7860-8452}
\email{zhouhang2991@gmail.com}
\affiliation{%
  \institution{Simon Fraser University}
  \city{Burnaby}
  \state{British Columbia}
  \country{Canada}
}

\author{Weiming Zhang}
\orcid{0000-0001-5576-6108}
\email{zhangwm@ustc.edu.cn}
\affiliation{%
  \institution{University of Science and Technology of China}
  \city{Hefei}
  \country{China}
}

\begin{abstract}

In the past few years, the widespread use of 3D printing technology enables the growth of the market of 3D printed products. On Esty, a website focused on handmade items, hundreds of individual entrepreneurs are selling their 3D printed products. Inspired by the positive effects of machine-readable tags, like barcodes, on daily product marketing, we propose AnisoTag, a novel tagging method to encode data on the 2D surface of 3D printed objects based on \emph{reflection anisotropy}. AnisoTag has an unobtrusive appearance and much lower extraction computational complexity, contributing to a lightweight low-cost tagging system for individual entrepreneurs. On AnisoTag, data are encoded by the proposed tool as reflective anisotropic microstructures, which would reflect distinct illumination patterns when irradiating by collimated laser. Based on it, we implement a real-time detection prototype with inexpensive hardware to determine the reflected illumination pattern and decode data according to their mapping. We evaluate AnisoTag with various 3D printer brands, filaments, and printing parameters, demonstrating its superior usability, accessibility, and reliability for practical usage.

\end{abstract}




\begin{teaserfigure}
\setlength{\belowcaptionskip}{0pt}
\setlength{\abovecaptionskip}{0pt}
\begin{center}
  \includegraphics[width=\linewidth]{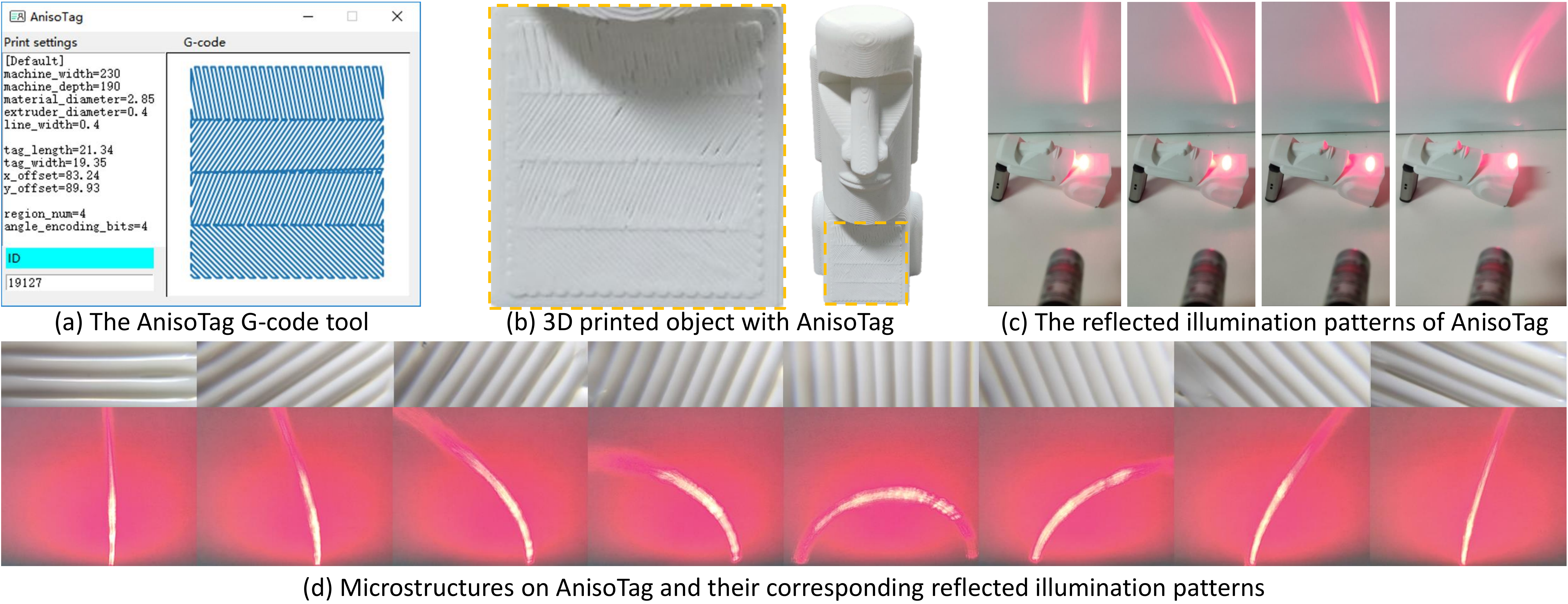}
  \caption{\textbf{Our proposed AnisoTag encodes data with smooth cylinder surface (SCS) microstructures.} (a) The \texttt{G-code} file of AnisoTag used for 3D printing is generated by the proposed \texttt{G-code} tool. (b) One example of AnisoTag is on the 2D surface of one 3D printed object. (c) When AnisoTag is irradiated by a collimated laser beam, its SCS microstructures with different angles would reflect obviously different illumination patterns, which we call \emph{reflection anisotropy}. (d) The micro view of one testing AnisoTag (top row) and the corresponding reflected illumination patterns (bottom row), which can be unidirectionally mapped to the embedded data for extraction.}
  \Description{Subfigure (a) shows a UI interface with the title AnisoTag, which has two text boxes on the left side of the interface, one for entering the parameters of 3D printing and the other for entering the embedded id. on the right side is a preview of the path of the G-code corresponding to the AnisoTag, consisting of many infilling lines. Subfigure (b) shows a 3D printed Moai figurine. It has a 2D surface with AnisoTag embedding on it. In subfigure (c), the Moai figurine is placed on a table. One collimated laser light generated from a laser pointer is shone in the AnisoTag on the Moai figurine, and the reflected light forms a clear curved illumination pattern on a white paper perpendicular to the table. The top row in subfigure (d) shows a series of 3D printed microstructures presented as smooth cylindrical sides arranged in parallel and having different angles from left to right. The bottom row shows the illumination patterns reflected from the corresponding microstructures, exhibiting as rotating curves.
  }
  \label{fig:teaser}
\end{center}
\end{teaserfigure}

\maketitle

\input{sections/introduction}

\input{sections/related_work}

\input{sections/fabrication}

\input{sections/detection}

\input{sections/experiments}

\begin{acks}
We thank CHI23's anonymous reviewers for their constructive feedback that helped improve the paper. We thank Chenxi Sun for the help in the prototype implementation. This work was supported in part by the Natural Science Foundation of China under Grant 62072421, 62002334, 62102386, 62121002, and U20B2047.
\end{acks}

\balance
\bibliographystyle{ACM-Reference-Format}
\bibliography{main}

\appendix
\input{sections/appendix.tex}

\end{document}

%% file: sections/introduction.tex
\aptLtoX[graphic=no,type=env]{
\begin{table*}[t]
\label{tab:comparison}
\scalebox{.95}{
\begin{tabular}{lccccccc}
\hline
Method & Invisibility & \begin{tabular}[c]{@{}c@{}}Availability \\ in complex shapes\end{tabular} & \begin{tabular}[c]{@{}c@{}}All-in-one\\ fabrication\end{tabular} & Fabrication device & Decoding device & Real-time extraction \\ \hline

Barcode & \cellcolor[HTML]{ED7D31}-\,- & \cellcolor[HTML]{FFE699} \XSolidBrush & \cellcolor[HTML]{ED7D31} \XSolidBrush& \cellcolor[HTML]{99CC66} Paper printer & \cellcolor[HTML]{99CC66}RGB camera & \cellcolor[HTML]{99CC66} \Checkmark \\ \hline

RFID tags \cite{RFID2,RFID3} & \cellcolor[HTML]{99CC66}++ & \cellcolor[HTML]{99CC66} \Checkmark & \cellcolor[HTML]{ED7D31} \XSolidBrush& \cellcolor[HTML]{ED7D31}Insertion devices & \cellcolor[HTML]{99CC66}RFID reader & \cellcolor[HTML]{99CC66} \Checkmark \\ \hline

Hou et al. \cite{layer_artifact} & \cellcolor[HTML]{99CC66}+ & \cellcolor[HTML]{99CC66} \Checkmark & \cellcolor[HTML]{99CC66} \Checkmark & \cellcolor[HTML]{99CC66}FDM & \cellcolor[HTML]{FFE699}3D scanner & \cellcolor[HTML]{ED7D31} \XSolidBrush\\ \hline

Delmotte et al. \cite{layer_thickness} & \cellcolor[HTML]{99CC66}+ & \cellcolor[HTML]{FFE699} \XSolidBrush & \cellcolor[HTML]{99CC66} \Checkmark & \cellcolor[HTML]{99CC66}FDM & \cellcolor[HTML]{99CC66}Paper scanner & \cellcolor[HTML]{ED7D31} \XSolidBrush\\ \hline

AirCode \cite{aircode} & \cellcolor[HTML]{99CC66}++ & \cellcolor[HTML]{FFE699} \XSolidBrush& \cellcolor[HTML]{99CC66} \Checkmark & \cellcolor[HTML]{FFE699}SLA or similar & \cellcolor[HTML]{FFE699}Projector  + camera & \cellcolor[HTML]{ED7D31} \XSolidBrush\\ \hline

InfraStructs \cite{InfraStructs} & \cellcolor[HTML]{99CC66}++ & \cellcolor[HTML]{FFE699} \XSolidBrush& \cellcolor[HTML]{99CC66} \Checkmark & \cellcolor[HTML]{FFE699}SLA or similar & \cellcolor[HTML]{ED7D31}Terahertz camera & \cellcolor[HTML]{ED7D31} \XSolidBrush \\ \hline

Acoustic Voxels \cite{Acoustic_Voxels} & \cellcolor[HTML]{FFE699}+/- & \cellcolor[HTML]{99CC66} \Checkmark & \cellcolor[HTML]{99CC66} \Checkmark & \cellcolor[HTML]{99CC66}All & \cellcolor[HTML]{99CC66}Speaker +  microphone & \cellcolor[HTML]{99CC66} \Checkmark \\ \hline

Acoustic barcodes \cite{acoustic_barcodes} & \cellcolor[HTML]{ED7D31}-\,- & \cellcolor[HTML]{FFE699} \XSolidBrush & \cellcolor[HTML]{99CC66} \Checkmark & \cellcolor[HTML]{99CC66}All & \cellcolor[HTML]{99CC66}microphone & \cellcolor[HTML]{99CC66} \Checkmark \\ \hline

G-ID \cite{G-id} (bottom) & \cellcolor[HTML]{99CC66}++ & \cellcolor[HTML]{FFE699} \XSolidBrush & \cellcolor[HTML]{99CC66} \Checkmark & \cellcolor[HTML]{99CC66}FDM & \cellcolor[HTML]{99CC66}RGB camera & \cellcolor[HTML]{ED7D31} \XSolidBrush\\ \hline

G-ID (infill) & \cellcolor[HTML]{99CC66}++ & \cellcolor[HTML]{FFE699} \XSolidBrush & \cellcolor[HTML]{99CC66} \Checkmark & \cellcolor[HTML]{99CC66}FDM & \cellcolor[HTML]{99CC66}Light source  + camera & \cellcolor[HTML]{ED7D31} \XSolidBrush\\ \hline

LayerCode \cite{layercode} (color) & \cellcolor[HTML]{ED7D31}-\,- & \cellcolor[HTML]{99CC66} \Checkmark & \cellcolor[HTML]{99CC66} \Checkmark & \cellcolor[HTML]{ED7D31}Dual color FDM & \cellcolor[HTML]{99CC66}RGB camera & \cellcolor[HTML]{ED7D31} \XSolidBrush\\ \hline

LayerCode (near-infrared) & \cellcolor[HTML]{99CC66}++ & \cellcolor[HTML]{99CC66} \Checkmark & \cellcolor[HTML]{99CC66} \Checkmark & \cellcolor[HTML]{ED7D31}Dual color SLA & \cellcolor[HTML]{FFE699}Camera +  NIR filter & \cellcolor[HTML]{ED7D31} \XSolidBrush\\ \hline

LayerCode (height) & \cellcolor[HTML]{FFE699}+/- & \cellcolor[HTML]{99CC66} \Checkmark & \cellcolor[HTML]{99CC66} \Checkmark & \cellcolor[HTML]{99CC66}FDM & \cellcolor[HTML]{99CC66}RGB camera & \cellcolor[HTML]{ED7D31} \XSolidBrush\\ \hline 
\hline

AnisoTag & \cellcolor[HTML]{99CC66}+ & \cellcolor[HTML]{FFE699} \XSolidBrush & \cellcolor[HTML]{99CC66} \Checkmark & \cellcolor[HTML]{99CC66}FDM & \cellcolor[HTML]{99CC66}Detection  prototype & \cellcolor[HTML]{99CC66} \Checkmark \\ \hline

AnisoTag (camera) $^*$ & \cellcolor[HTML]{99CC66}+ & \cellcolor[HTML]{FFE699} \XSolidBrush & \cellcolor[HTML]{99CC66} \Checkmark & \cellcolor[HTML]{99CC66}FDM & \cellcolor[HTML]{99CC66}RGB camera & \cellcolor[HTML]{ED7D31} \XSolidBrush\\ \hline
\end{tabular}
}
\setlength{\belowcaptionskip}{3pt}
\setlength{\abovecaptionskip}{0pt}
\caption{\textbf{Comparison with related work that encodes data on 3D  printed objects.} The information capacity of the above methods ranges from a few bits to over 64 bits. Compared with these methods, our proposed AnisoTag has a middle information capacity of around 50 bits when set as the same size as one credit card. And its capacity can be extended to 160 bits (Section~\ref{sec:highcapacitytag}) if it is extracted by an RGB camera and image processing operations. To the artifacts generated in the data encoding process, its invisibility would be classified as `++' if it is completely invisible to the human eye; `+' if it is unobtrusive unless viewed closely; `+/-' if it could be noticed at first glance; `- -' if it is visually obtrusive.  To the devices for fabrication and decoding, their background color represents the popularity of the device: green represents consumer-grade devices and red represents expensive ones.}
\Description{
AnisoCard is the only scheme in the table that embeds data on 3D printed objects with consumer-grade 3D printers, has comparable information capacity, and can be extracted in real time with inexpensive hardware.
}
\end{table*}}
{
\begin{table*}[t]
\label{tab:comparison}
\scalebox{.95}{
\begin{tabular}{lccccccc}
\hline
Method & Invisibility & \begin{tabular}[c]{@{}c@{}}Availability \\ in complex shapes\end{tabular} & \begin{tabular}[c]{@{}c@{}}All-in-one\\ fabrication\end{tabular} & Fabrication device & Decoding device & Real-time extraction \\ \hline

Barcode & \cellcolor[HTML]{ED7D31}-\,- & \cellcolor[HTML]{FFE699} \XSolidBrush & \cellcolor[HTML]{ED7D31} \XSolidBrush& \cellcolor[HTML]{99CC66} Paper printer & \cellcolor[HTML]{99CC66}RGB camera & \cellcolor[HTML]{99CC66} \Checkmark \\ \hline

RFID tags \cite{RFID2,RFID3} & \cellcolor[HTML]{99CC66}++ & \cellcolor[HTML]{99CC66} \Checkmark & \cellcolor[HTML]{ED7D31} \XSolidBrush& \cellcolor[HTML]{ED7D31}Insertion devices & \cellcolor[HTML]{99CC66}RFID reader & \cellcolor[HTML]{99CC66} \Checkmark \\ \hline

Hou et al. \cite{layer_artifact} & \cellcolor[HTML]{99CC66}+ & \cellcolor[HTML]{99CC66} \Checkmark & \cellcolor[HTML]{99CC66} \Checkmark & \cellcolor[HTML]{99CC66}FDM & \cellcolor[HTML]{FFE699}3D scanner & \cellcolor[HTML]{ED7D31} \XSolidBrush\\ \hline

Delmotte et al. \cite{layer_thickness} & \cellcolor[HTML]{99CC66}+ & \cellcolor[HTML]{FFE699} \XSolidBrush & \cellcolor[HTML]{99CC66} \Checkmark & \cellcolor[HTML]{99CC66}FDM & \cellcolor[HTML]{99CC66}Paper scanner & \cellcolor[HTML]{ED7D31} \XSolidBrush\\ \hline

AirCode \cite{aircode} & \cellcolor[HTML]{99CC66}++ & \cellcolor[HTML]{FFE699} \XSolidBrush& \cellcolor[HTML]{99CC66} \Checkmark & \cellcolor[HTML]{FFE699}SLA or similar & \cellcolor[HTML]{FFE699}\begin{tabular}[c]{@{}c@{}}Projector \\ + camera\end{tabular} & \cellcolor[HTML]{ED7D31} \XSolidBrush\\ \hline

InfraStructs \cite{InfraStructs} & \cellcolor[HTML]{99CC66}++ & \cellcolor[HTML]{FFE699} \XSolidBrush& \cellcolor[HTML]{99CC66} \Checkmark & \cellcolor[HTML]{FFE699}SLA or similar & \cellcolor[HTML]{ED7D31}Terahertz camera & \cellcolor[HTML]{ED7D31} \XSolidBrush \\ \hline

Acoustic Voxels \cite{Acoustic_Voxels} & \cellcolor[HTML]{FFE699}+/- & \cellcolor[HTML]{99CC66} \Checkmark & \cellcolor[HTML]{99CC66} \Checkmark & \cellcolor[HTML]{99CC66}All & \cellcolor[HTML]{99CC66}\begin{tabular}[c]{@{}c@{}}Speaker + \\ microphone\end{tabular} & \cellcolor[HTML]{99CC66} \Checkmark \\ \hline

Acoustic barcodes \cite{acoustic_barcodes} & \cellcolor[HTML]{ED7D31}-\,- & \cellcolor[HTML]{FFE699} \XSolidBrush & \cellcolor[HTML]{99CC66} \Checkmark & \cellcolor[HTML]{99CC66}All & \cellcolor[HTML]{99CC66}microphone & \cellcolor[HTML]{99CC66} \Checkmark \\ \hline

G-ID \cite{G-id} (bottom) & \cellcolor[HTML]{99CC66}++ & \cellcolor[HTML]{FFE699} \XSolidBrush & \cellcolor[HTML]{99CC66} \Checkmark & \cellcolor[HTML]{99CC66}FDM & \cellcolor[HTML]{99CC66}RGB camera & \cellcolor[HTML]{ED7D31} \XSolidBrush\\ \hline

G-ID (infill) & \cellcolor[HTML]{99CC66}++ & \cellcolor[HTML]{FFE699} \XSolidBrush & \cellcolor[HTML]{99CC66} \Checkmark & \cellcolor[HTML]{99CC66}FDM & \cellcolor[HTML]{99CC66}\begin{tabular}[c]{@{}c@{}}Light source \\ + camera\end{tabular} & \cellcolor[HTML]{ED7D31} \XSolidBrush\\ \hline

LayerCode \cite{layercode} (color) & \cellcolor[HTML]{ED7D31}-\,- & \cellcolor[HTML]{99CC66} \Checkmark & \cellcolor[HTML]{99CC66} \Checkmark & \cellcolor[HTML]{ED7D31}Dual color FDM & \cellcolor[HTML]{99CC66}RGB camera & \cellcolor[HTML]{ED7D31} \XSolidBrush\\ \hline

LayerCode (near-infrared) & \cellcolor[HTML]{99CC66}++ & \cellcolor[HTML]{99CC66} \Checkmark & \cellcolor[HTML]{99CC66} \Checkmark & \cellcolor[HTML]{ED7D31}Dual color SLA & \cellcolor[HTML]{FFE699}\begin{tabular}[c]{@{}c@{}}Camera + \\ NIR filter\end{tabular} & \cellcolor[HTML]{ED7D31} \XSolidBrush\\ \hline

LayerCode (height) & \cellcolor[HTML]{FFE699}+/- & \cellcolor[HTML]{99CC66} \Checkmark & \cellcolor[HTML]{99CC66} \Checkmark & \cellcolor[HTML]{99CC66}FDM & \cellcolor[HTML]{99CC66}RGB camera & \cellcolor[HTML]{ED7D31} \XSolidBrush\\ \hline 
\hline

AnisoTag & \cellcolor[HTML]{99CC66}+ & \cellcolor[HTML]{FFE699} \XSolidBrush & \cellcolor[HTML]{99CC66} \Checkmark & \cellcolor[HTML]{99CC66}FDM & \cellcolor[HTML]{99CC66}\begin{tabular}[c]{@{}c@{}}Detection \\ prototype\end{tabular} & \cellcolor[HTML]{99CC66} \Checkmark \\ \hline

AnisoTag (camera) $^*$ & \cellcolor[HTML]{99CC66}+ & \cellcolor[HTML]{FFE699} \XSolidBrush & \cellcolor[HTML]{99CC66} \Checkmark & \cellcolor[HTML]{99CC66}FDM & \cellcolor[HTML]{99CC66}RGB camera & \cellcolor[HTML]{ED7D31} \XSolidBrush\\ \hline
\end{tabular}
}
\setlength{\belowcaptionskip}{3pt}
\setlength{\abovecaptionskip}{0pt}
\caption{\textbf{Comparison with related work that encodes data on 3D  printed objects.} The information capacity of the above methods ranges from a few bits to over 64 bits. Compared with these methods, our proposed AnisoTag has a middle information capacity of around 50 bits when set as the same size as one credit card. And its capacity can be extended to 160 bits (Section~\ref{sec:highcapacitytag}) if it is extracted by an RGB camera and image processing operations. To the artifacts generated in the data encoding process, its invisibility would be classified as `++' if it is completely invisible to the human eye; `+' if it is unobtrusive unless viewed closely; `+/-' if it could be noticed at first glance; `- -' if it is visually obtrusive.  To the devices for fabrication and decoding, their background color represents the popularity of the device: green represents consumer-grade devices and red represents expensive ones.}
\Description{
AnisoCard is the only scheme in the table that embeds data on 3D printed objects with consumer-grade 3D printers, has comparable information capacity, and can be extracted in real time with inexpensive hardware.
}
\end{table*}
}

\section{Introduction}
With the popularization of 3D printing technology and the improvement of the print quality of consumer-grade printers, entrepreneurs have been able to develop consumer-quality products without the traditional expense of mass manufacturing \cite{smith2019developing}, which allowed for the growth of small businesses and individual digital maker-entrepreneurs. On websites such as Amazon and Etsy \cite{etsy}, thousands of 3D printed products, e.g., figurines, cosplay clothing, and useful household items, are being sold. Besides online eCommerce platforms, digital maker-entrepreneurs sold their 3D printed products in consignment shops and local groceries \cite{smith2019developing,troxler2017digital,petersen2017impact}.

In the past decades, machine-readable tags are used in a wide range of industries, from manufacturing and logistics to mobile marketing and business \cite{device}. A representative use case is the barcode, a technological staple in inventory and sales procedures. We believe that a machine-readable tag would considerably benefit the market of 3D printed products, just like barcodes for everyday products. For example, through the tag embedded in one 3D printed product, people could easily access the author's website on Etsy for similar products, learn the story of the product design, and know the price, which is the most important for purchasers. Moreover, fabricating that tag with the existing 3D printer rather than post-processing operations is an attractive idea, meaning no extra device or time cost for individual entrepreneurs.

When designing a tagging scheme for 3D printed products, we focus on several important properties. (1) The entire system should bring as less extra cost as possible for easier implementation. In an ideal case, this machine-readable tag is only fabricated by the existing 3D printer of sellers. (2) The tag should be unobtrusive. (3) The tag should ideally be detectable in real time to pursue a similar application to the barcode. Although there are plenty of schemes to embed data on 3D printed objects having been proposed, such as AirCode, LayerCode, G-ID, to name a few, the existing schemes hardly meet the above demands. AirCode \cite{aircode} is a carefully designed 2D barcode embedded as air pockets beneath the surface of the 3D printed semitransparent object for object tagging. Yet its fabrication requires high-precision 3D printers, and its detection requires a projector and camera setup; resulting in limited implementation in everyday scenarios. Objects embedded with LayerCode \cite{layercode} are printed by two-type 3D printing filaments, e.g., different colors, heights, or reflections under near-infrared light, to encode data. G-ID \cite{G-id} encodes data by the infilling lines in the 3D printing process, including different widths and angles. Both of LayerCode and G-ID are decoded through the camera and computationally complex image processing approaches, hard to meet the real-time requirement of barcode applications.

In this paper, we propose a machine-readable tag for 3D printed products, named as AnisoTag. It is a lightweight, inexpensive, and easy-to-implement tagging system on 3D printed products. Specifically, according to \cite{anisotropic,anisotropic_model,anisotropic_structure}, the characteristics of anisotropic reflectance of one material are directly determined by its surface microstructures. We leverage the FDM 3D printing process to fabricate unobtrusive microstructures on AnisoTag to embed data and make it reflective anisotropic. Inspired by surface inspection \cite{clarke1985panel,lee1999analysis,huynh1993an}, the illumination pattern reflected by reflective anisotropic AnisoTag is leveraged to determine the corresponding microstructure and the embedded data. Experimentally we validate the functionality of our AnisoTag with various printing settings including different 3D printer brands, materials, and 3D printing parameters.  

The main contributions of this paper are summarized as follows:
\begin{itemize}
    \item A unobtrusive machine-readable tag on the 2D surface of 3D printed objects, named as AnisoTag, which has low requirements for devices: an entry-level FDM 3D printer and consumer-grade filaments. Thus, individual entrepreneurs could place hyperlinks to their 3D printed products for sales management or advertising delivery.
    \item A lightweight data extraction method exploiting the reflection anisotropy of AnisoTag with a much lower computational cost compared to the image processing algorithm. Targeting an extractor similar to the handle barcode scanner, we design a self-contained detection prototype with inexpensive hardware, in which the collimated laser is adopted as the lighting source to maximize the utilization of reflection anisotropy.
    \item A tool for users to directly generate the \texttt{G-code} file\footnote{\texttt{G-code} file: The input file of 3D printer, instructing it to fabricate the corresponding objects.} of AnisoTag with adjustable printing parameters to suit diverse scenarios and 3D printer types.
\end{itemize}

%% file: sections/related_work.tex
\section{Related Work}
\label{related_work}
We cover related work on copyright protection, human-computer interaction of 3D printed objects, as well as 3D printed barcodes.

\subsection{Copyright Protection}
With the development of the 3D printing content industry, copyright issues may occur in 3D mesh and 3D printed objects, and the corresponding watermarking methods are proposed to protect copyright. Hou et al. \cite{layer_hist} propose a non-blind 3D mesh watermarking method using a histogram-based circular shift coding structure, which is 3D print-scan resilient. They further propose a blind watermarking method leveraging layering artifacts in \cite{layer_artifact}. Peng et al. \cite{printed_signature} build a 3D printed object authentication framework composed of registration and verification processes. When 3D printing one object, a special authentication mark would be fabricated on it, whose printing noise would be registered. In the verification stage, the 3D printed object is authenticated by comparing the extracted printing noise with the registered one. These methods embed watermarks in digital 3D mesh before the 3D printing process. Additionally, the method proposed by Delmotte et al. \cite{layer_thickness} embeds a blind watermark by modifying layer thickness in the \texttt{G-code} file. Song et al. \cite{Genuine_QR_Codes} propose using a 3D printer fingerprint generated by its hardware characteristics to prevent counterfeiting, without modifying digital meshes or \texttt{G-code} files. As forensics methods, these methods need large equipment, such as a high-precision 3D scanner and document scanner, to extract the embedded watermark data, which prevents these solutions from being used in retail scenarios.

\subsection{Human-Computer Interaction}
\label{sec:HCI}
Some methods embed tags on 3D printed objects to build human-computer hyperlinks with human-invisible modifications. Although the existing well-proven methods, such as RFID chips \cite{RFID1,RFID2,RFID3} and optical barcodes \cite{device}, could also build hyperlinks, considering additional fabrication costs and demands, this category of methods focuses more on fabricating both the 3D object and hyperlinking tags in one 3D printing process and has yielded some promising applications.

AirCode \cite{aircode} embeds information as air pockets beneath a 3D printed semitransparent object, which is human-imperceptible under a daily light environment, whose encoded data can be read by computational imaging approaches using a telephoto camera and a projector. Willis and Wilson \cite{InfraStructs}~present InfraStructs, in which information is embedded as the encoded structure inside 3D printed objects. And a Terahertz imaging system is further utilized to read the inside information, as Terahertz can penetrate through most materials used in 3D printing. AirCode and InfraStructs can be applied for pose estimation, robotic grasping, and object identification. Acoustic Voxels \cite{Acoustic_Voxels} can automate the design of acoustic filters with complex geometries. By optimizing its internal structure, tags can be embedded into the acoustic filtering effects of a shape. In InfraredTags \cite{infraredTags}, 3D objects are printed with infrared transmitting filaments, and then data is embedded with common material or air gaps into these 3D printed objects. The above methods require no electronics or multi-material during fabrication, yet they all require high-precision 3D printers to produce internal structures, which inevitably increases the fabrication cost. Additionally, most of them need specialized equipment and a series of image processing operations for extraction, having relatively high demands on the accuracy and computing power of extraction equipment.

Besides embedding information inner the 3D printed objects, there have been some methods that directly tag information on the surface of 3D printed objects using unobtrusive characteristics. G-ID \cite{G-id} tags the 3D printed object with the angle and thickness of infill patterns generated naturally in the 3D printing process. The outline of the embedded plane would be registered in the image library and used to align the camera-captured G-ID, which can be read by image processing approaches to extract the encoded infill patterns. ElSayed et al. \cite{speed_control} discover that the 3D printing speed variations would introduce subtle localized height differences on the surface of objects, and proposes a corresponding information embedding method. It utilizes an optical profilometer to predict the embedded data for each embedding region on the surface of the object. Both of them could be fabricated by consumer-grade 3D printers. Still, their extraction process involves complex image processing, which is difficult to achieve in real time on devices with low computational power.

Considering that the linewidth of infill lines in 3D printing is usually set as $0.4mm$, it is quite easy to generate microstructures that are easily overlooked by human eyes. But this also means these microstructures cannot be directly distinguished by common image sensors. Therefore, how to design unobtrusive microstructures to embed data while the embedded data can be efficiently read with simple hardware and extraction algorithms is worth investigating. 

\subsection{Barcode}
Researchers embed optical tags, like barcodes or QR codes, on 3D printed objects to establish hyperlinks between physical objects and digital content. For 3D printed objects, compared to attaching barcodes to them with stickers or inkjet that are easily worn, printing 3D objects along with stable physical barcodes is a better choice. Harrison et al. \cite{acoustic_barcodes} present acoustic barcodes composed of a series of parallel grooves on a flat surface, which are easily fabricated by 3D printing. Swiping acoustic barcodes with a hard object can generate an encoded acoustic waveform, whose data could be extracted with a low-cost contact microphone. Maia et al. \cite{layercode} propose LayerCode, which is encoded in 3D printing layers by two different states of 3D printing materials and performs well on a non-planar surface of 3D printed objects. The data embedded by LayerCode can be further extracted by a customized image processing approach. Besides barcode, 3D printed QR code methods have been developed recently \cite{B-spline,self-shadows,Directional}. Kikuchi et al. \cite{B-spline} propose to utilize ambient occlusion generated by 3D printed grooves to represent dark modules in QR code, which can be implemented with any type of filaments. Peng et al. \cite{self-shadows} fabricate QR codes with minimal modification of 3D mesh shapes. They \cite{Directional} further improve the unobtrusiveness of 3D printed QR codes leveraging directional light. As optical 3D tagging methods, the above generally have perceptible artifacts to humans and their embedded data are extracted based on image processing approaches, which usually burden a certain computational complexity.

%% file: sections/fabrication.tex
\section{AnisoTag Fabrication}
The AnisoTag we pursue is an easy-to-implement machine-readable tag for 3D printed products. The data are encoded into the 3D printed surface microstructures and can be detected by their reflection anisotropy. In general, the reflection anisotropy of materials reflects the inconsistency of its appearance of different lighting and viewing direction conditions. For AnisoTag fabrication, we aim to ensure the easy-to-detect reflection anisotropy of 3D printed microstructures.

In this section, we first discuss why FDM 3D printing crafts microstructures on the surface of a 3D printed object that leads to reflection anisotropy. Then, we propose a microstructure type with obvious reflection anisotropy and mathematically establish a reflection model to illustrate the effectiveness of such a design. Finally, to implement that microstructure, we present a \texttt{G-code} tool to encode data into reflective anisotropic microstructures. Note that \texttt{G-code} files are the most widely used 3D printer instruction files.

\begin{figure}[t]
\setlength{\belowcaptionskip}{-8pt}
\setlength{\abovecaptionskip}{3pt}
  \begin{center}
  \includegraphics[width=3.3in]{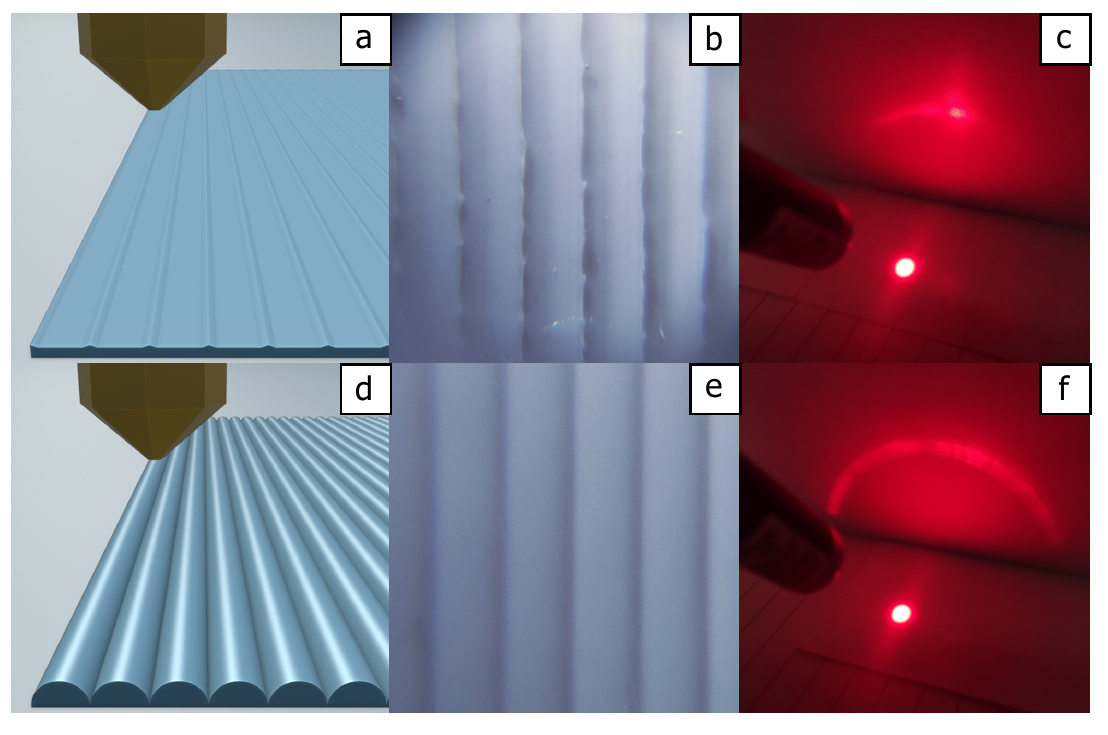}
  \caption{\textbf{Visualization of two printing settings.} (a) The schematic diagram of the filament extruding process under normal 3D printing settings. Each edge of the infilling line is usually warped because of the joint effects of the nozzle pressure and the mobility of the melting filament. (b) The actual microscopic view of that infilling line. (c) The reflected illumination pattern under a collimated laser beam from a laser pointer pen. (d) The schematic diagram of the SCS microstructures used in AnisoTag. (e)(f) The corresponding microscopic view and the reflected illumination pattern.}
  \Description{Visualization of two printing settings. The reflected illumination pattern in (c) is blurred and has no clear shape. (d)(e) are fabricated with modified settings, having SCS microstructures, which reflect a clear illumination pattern in (f), i.e., a clear curve line on the background plane.}
  \label{micro_design}
  \end{center}
\end{figure}

\begin{figure*}[t]
\setlength{\belowcaptionskip}{0pt}
\setlength{\abovecaptionskip}{3pt}
  \begin{center}
  \includegraphics[width=6in]{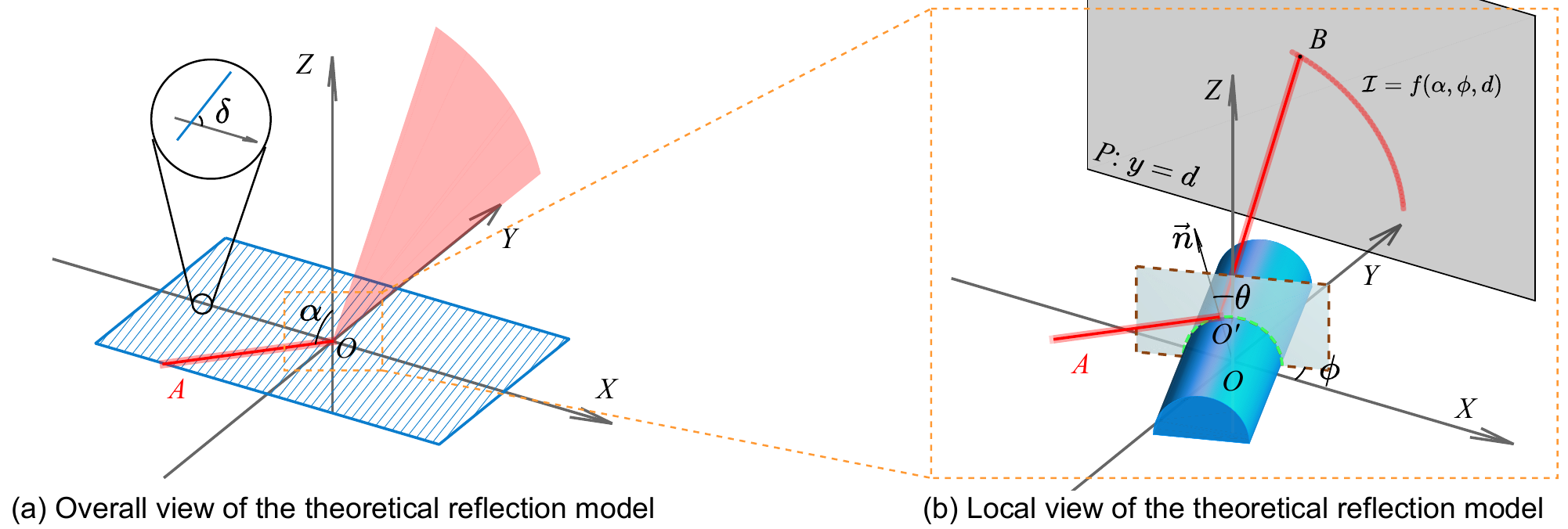}
  \caption{\textbf{Theoretical reflection model of 3D printed tags with SCS microstructures.} The presented 3D printed tag is infilled by SCS microstructures with the same cylindrical axis of angle $\delta$. In the general view (a), the red line along $\vec{AO}$ is the incident ray, whose direction is defined as $\vec{a}$. And the incident angle $\angle AOZ=\alpha$. In the microscopic view (b), the SCS microstructure is modeled as a half-cylinder with a cylindrical section angle $\phi=\delta-\pi/2$. $O^\prime$ is the point of incidence located on the surface of the SCS microstructure. And the surface normal of $O^\prime$ is along the direction of $\vec{OO^\prime}$, defined as $\vec{n}$. The angle between $\vec{n}$ and $Z$-axis is $\theta$. The reflected ray forms a reflected illumination pattern $\mathcal{I}$ on the background plane $P: y=d$.}
  \Description{The SCS microstructure is modeled as a half-cylinder, which reflects the laser beam as one illumination pattern, uniquely mapping to the arranging angle of SCS microstructures.}
  \label{coordinate_system}
  \end{center}
\end{figure*}

\subsection{Microstructure Analysis}
Fused Deposition Modeling (FDM) 3D printers are one of the most popular types with low-priced devices and printing materials \cite{hubs18,hubs20}, especially for individual entrepreneurs of 3D printed products \cite{kim2019study}. In an FDM system, the printing material, usually a thermoplastic filament, is melted and extruded by a heated printer nozzle. The \texttt{G-code} file, as the input of the 3D printer, controls the trajectory of nozzle and extrusion amount to print one object layer by layer, where each layer is infilled by the melt printing material along with a preset trajectory. Thus, the surface of the 3D printed objects generally retains microstructures fabricated by the melted material and the nozzle of the printer; see Figure~\ref{micro_design}(b).

As discussed in Section~\ref{sec:HCI}, some previous methods utilize 3D printed microstructures to embed data to establish human-computer hyperlinks. To analyze these microstructures, they usually employ a camera or a scanner, supplemented with signal processing operations. Yet tagging methods with better visual quality, i.e., more unobtrusive microstructures, commonly require higher-precision equipment and more complex extraction algorithms, which are not easy to implement for individual entrepreneurs with entry-level equipment. Thus, to design an easy-to-implement machine-readable tag system, we desire a lightweight microstructure detection method. Inspired by surface inspection methods based on laser beam \cite{clarke1985panel,lee1999analysis,huynh1993an}, where a laser beam is used, we believe that the optical properties of 3D printed objects, i.e. reflection anisotropy, could be leveraged to design a simple but effective microstructure detection method, in which the reflection anisotropy could be detected by the collimated laser beam generated by one inexpensive laser pointer pen.

According to the Ashikhmin-Shirley model \cite{anisotropic_model}, any nonhomogeneous microstructures, including the one shown in Figure~\ref{micro_design}(b), would induce the reflection anisotropy of the material. However, as shown in Figure~\ref{micro_design}(c), the reflection anisotropy generated by regular 3D printing is too fragile to be utilized for robust microstructure detection, especially in the case of low-cost detection. Thus, we aim to design an effective microstructure type with easy-to-detect reflection anisotropy for fabricating AnisoTag.

\subsection{SCS Microstructure}
\label{anisotropy_analysis}
The design of our target microstructure is driven by two key goals. The first is to preserve surface smoothness, ensuring effective reflection anisotropy under a collimated laser beam. The second is to preserve minimal modifications to the 3D printing process. To satisfy the above demands, we design the smooth cylinder surface (SCS) microstructure to generate an effective ray reflection effect; see Figure~\ref{micro_design}(d\textasciitilde f). The SCS microstructure follows the same 3D printing process, except that some infilling parameters are modified to fabricate smooth half-cylindrical infilling lines.

Below we analyze the reflection process on the SCS microstructure to illustrate its reflection anisotropy. Consider the scenario shown in Figure~\ref{coordinate_system}(a): In the Cartesian coordinate system, one 3D printed tag with SCS microstructure is placed on the $XoY$ plane with the short side perpendicular to the $X$-axis. A collimated laser beam with the incident lighting source $A$ irradiates the tag on the $YoZ$ and then is reflected and illuminates the background plane perpendicular to the $Y$-axis. The point of incidence is set as origin $O$, and the incident angle is $\angle AOZ=\alpha$. In this case, if the tag owns reflection anisotropy, then the SCS microstructure with different angles of the infilling lines would reflect different illumination patterns. The angle between the cylindrical section and $X$-axis is denoted as $\phi$, and the angle between the cylindrical axis and $X$-axis is $\delta = \phi + \pi /2$. To simplify the statement, we use the cylindrical section angle $\phi$ and the cylindrical axis angle $\delta$ to denote the above angles in the following sections.

To analyze the pattern reflected by SCS microstructure, we enlarge the coordinate region near the origin $O$ to obtain coordinates system shown in Figure~\ref{coordinate_system}(b) and consider one single SCS microstructure, which is modeled as a half-cylinder in Figure~\ref{coordinate_system}(b). The diameter of the presented half-cylinder of SCS microstructure is approximately the nozzle aperture of the 3D printer, which is much smaller than the incident beam diameter (approximately $0.2mm \ll 5mm$). Thus, incident rays in this coordinates system could be regarded as parallel rays with the same incident angle $\alpha$ and the same incident direction $\vec{a}$. In Figure~\ref{coordinate_system}(b), we discuss the case that one of the parallel incident rays irradiates on the point of incidence $O^\prime$. 
Without loss of generality, we make the point of incidence $O^\prime$ and the origin $O$ at the same cylindrical section. Then, the surface normal $\vec{n}$ on the point of incidence $O^\prime$ can be determined by $\angle O^\prime OZ = \theta$. According to the law of reflection in the 3D coordinate system, the direction of the reflected ray $\vec{b}$ can be calculated by:
\begin{equation}
\label{reflected_vector}
    \begin{aligned}
    \vec{a}&=(0,\sin\alpha,-\cos\alpha),\\
    \vec{n}&=(\sin\theta \cos\phi,\sin\theta \sin\phi,\cos\theta),\\
    \vec{b}&=\vec{a}-2(\vec{n}\cdot\vec{a})\vec{n},
    \end{aligned}
\end{equation}
where $\vec{a}$ is the vector of the incident ray $\vec{AO^\prime}$ and $\vec{n}$ is the surface normal of $O^\prime$, having the same direction with $\vec{OO^\prime}$. To simplify the calculation, we approximate that the reflected ray is coming from the origin $O$ rather than the incident point $O^\prime$, as the distance between $O^\prime$ and the origin $O$ is much smaller than the distance between $O^\prime$ and the background plane $P: y=d$ (approximately $0.2 mm \ll 10 cm$). The reflected ray exits from origin $O$ and intersects the plane $P$ at the point $B$; see Figure~\ref{coordinate_system}(b). For the other points of incidence at the same cylindrical section with $O^\prime$, their position could be determined by $\theta \in (-\pi/2, \pi/2)$ and they reflect other illumination points on the background plane. Under the parallel incident rays, different illumination points on the background plane are reflected simultaneously and form the reflected illumination pattern $\mathcal{I}=f(\alpha, \phi, d)$, which is the function of incident angle $\alpha$, cylindrical section angle $\phi$, and the background plane location $d$. The specific derivation process of $\mathcal{I}=f(\alpha, \phi, d)$ is illustrated in the Appendix. The relationship between $\mathcal{I}$ and $\phi$ proves the reflection anisotropy of SCS microstructure, and the obvious reflected illumination patterns difference (Figure~\ref{micro_design}) further illustrates the validity of the SCS microstructure and the correctness of our model. In Section~\ref{anisotropy_analysis_2}, we further discuss the inferences of this model to guide the design of our microstructure detection prototype.

\begin{figure*}[t]
  \begin{center}
  \includegraphics[width=5.8in]{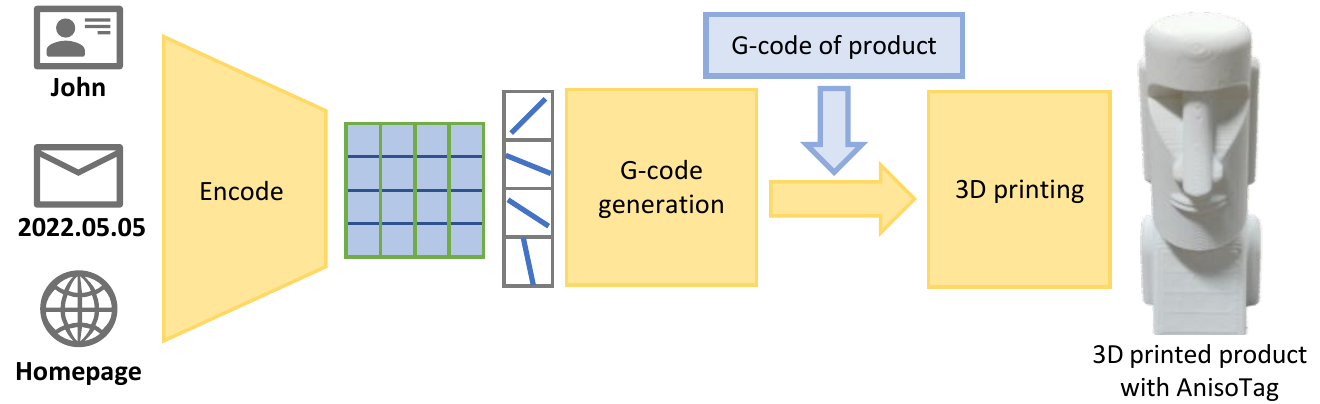}
  \caption{\textbf{Workflow of AnisoTag \texttt{G-code} tool and 3D printing process.} In this example, AnisoTag is divided into $n=4$ encoding regions and each one encodes $m=4$ bits, i.e., $16$ angles. The tool encodes the input data into a set of angles $\delta_i$ and then generates the AnisoTag \texttt{G-code} file based on these angles and the settings of the 3D printer. Finally, the \texttt{G-code} file of AnisoTag and the target model are combined to fabricate the 3D printed product with AnisoTag.}
  \Description{The input data, e.g., name, would be encoded as a set of angles, and then converted to the G-code file of AnisoTag by the proposed G-code tool. Then, the G-code of AnisoTag would be combined with the G-code of the target object to fabricate a 3D printed object with AnisoTag.}
  \label{gui_workflow}
  \end{center}
\end{figure*}

\subsection{Microstructure Implementation}
\label{GUI}
After determining the microstructure type utilized for AnisoTag, we can utilize the characteristics of the FDM printer and modify the FDM printing parameters, i.e. \texttt{G-code} file, to implement SCS microstructures, in which data is embedded. Then, the \texttt{G-code} of AnisoTag would be added into the \texttt{G-code} of the 3D printed product to fabricate the tagged product. The workflow of this process is shown in Figure~\ref{gui_workflow}. And we design a matching \texttt{G-code} tool to fabricate AnisoTag, which is illustrated as follows.

\subsubsection{AnisoTag setting}
\label{AnisoTag_setting}
To fabricate the target AnisoTag, some settings should be loaded into the \texttt{G-code} tool, e.g., the parameters of the involved 3D printer, the target size of AnisoTag, and some encoding settings. Different from existing 3D model slicing software that can only generate \texttt{G-code} file infilling the 3D printed layer with the same lines, our proposed \texttt{G-code} tool makes it possible to divide the 3D printed layer into several regions and apply different 3D printing parameters to each region. Thus, we could encode more information on AnisoTag under equivalent conditions. Specifically, we divide AnisoTag into several encoding regions along the parallel direction and each region has an independent infilling angle; see Figure~\ref{gui_workflow}. These infilling angles are determined by the embedded information. Under the reflection model in Figure~\ref{coordinate_system}, this partitioning method allows us to observe the illumination pattern reflected by different encoding regions by swiping AnisoTag along with the $X$-direction. The encoding settings consist of the number of encoding regions $n$ and the encoded bits $m$ of each region. The two parameters jointly determine the information capacity ($m \times n$ bits) of our AnisoTag.

\subsubsection{Data encoding}
\label{encoding}
The workflow of our proposed AnisoTag \texttt{G-code} tool consists of two steps: encoding input data into encoded infilling angles and then converting these angles to a \texttt{G-code} file. In this subsection, we provide details of how we encode input data into the infilling angles. The data is first converted into a bitstream and then encoded by error-correcting codes and zero-padded into a predefined length. It should be noted that to directly evaluate the data extraction performance of AnisoTag, we omit the error-correcting coding process in AnisoTag fabrication in this paper. The encoded bitstream is defined as $s$ with length $n \times m$, where $n$ is the number of encoding regions and $m$ is the number of encoded bits of each encoding region. To embed the bitstream $s$ into $n$ regions, we divide $s$ into non-overlapping $m$-bit sequence $s_i$, $i=1,2,...,n$. Each $s_i$ is treated as an $m$-bit gray code \cite{gray_code_wiki} to obtain its gray-code decimal value $v_i$. Taking the case of $m=2$ as an example, $2$-bit sequences $(00,01,10,11)_{Gray}$ would be converted to decimal values $(0,1,3,2)_{10}$. Gray coding makes the binary representation of adjacent values have only one-bit difference, which greatly decreases the bit extraction error rate because of the misjudgment of adjacent angles. Then, according to the value $v_i$ of $s_i$, the encoding region $r_i$, $i=1,2,...,n$, would be infilled by SCS microstructure with angle $\delta_i$. The mapping relationship between $\delta_i$ and $v_i$ is designed as follows:
\begin{equation}
\label{angle_mapping}
    \delta_i = \mathcal{M}(v_i/2^m \cdot 180^\circ), 
\end{equation}
where $\mathcal{M}(\cdot)$ is a nonlinear mapping function remapping the angular intervals to reflect more easily detectable illumination patterns with similar intervals. The specific definition of $\mathcal{M}(\cdot)$ is illustrated in the Appendix.

\subsubsection{G-code generation}
This subsection presents the second step of the AnisoTag \texttt{G-code} tool: converting the encoded infilling angles $\delta_i$ calculated from the above into the \texttt{G-code} file of AnisoTag. The \texttt{G-code} file includes a set of instructions for controlling a 3D printer. For example, the instruction format is usually shown as `\texttt{G1 X0.1 Y200 Z0.3 F1500 E15}', which means that it controls the nozzle of the 3D printer to move from the current position to coordinates (\texttt{X}, \texttt{Y}, \texttt{Z}): $(0.1, 200, 0.3)$ with speed \texttt{F}: $1500$ millimeters per minute and extrude \texttt{E}: $15mm$ filaments, guiding the 3D printer to fabricate one infilling line. Therefore, to generate the \texttt{G-code} file of AnisoTag, we need to make a slight modification to calculate all the infilling lines and convert them to instructions as below, mainly the calculation of \texttt{XYZE} values.

\begin{figure}[t]
  \begin{center}
  \includegraphics[width=\linewidth]{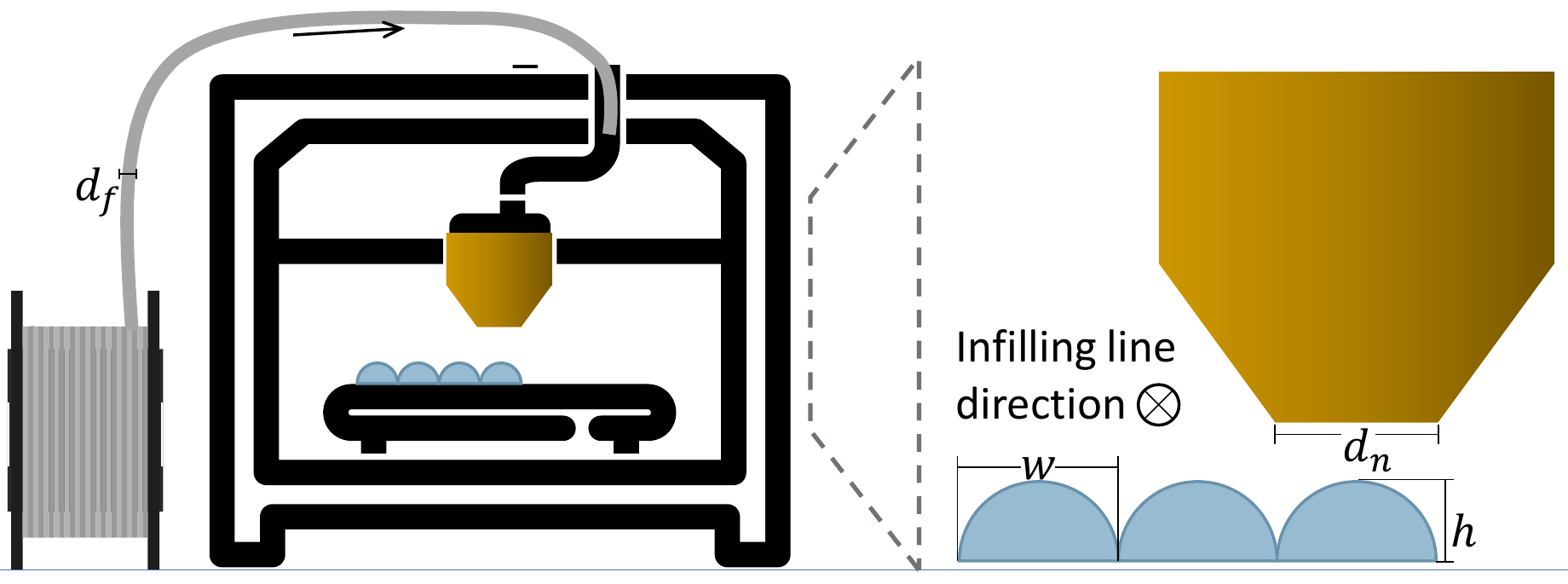}
  \caption{\textbf{Schematic diagram of the FDM 3D printing process.} Some 3D printer settings are marked: the filament diameter $d_f$, the nozzle diameter $d_n$, the linewidth $w$, and the layer height $h$. The filament is fed into the 3D printer, then melted and extruded as the infilling lines of the 3D printed object, whose printing direction is perpendicular to the plane of this figure. The right side of the figure shows the enlarged view of the extruder and the cross-section of infilling lines.}
  \Description{The filament is fed into the 3D printer, then melted and extruded as the infilling lines of the 3D printed object.}
  \label{gcode_cal}
  \end{center}
\end{figure}

First, based on the number of encoding regions and the size setting of AnisoTag, we obtain a set of $XY$ coordinates for the boundary of encoding regions. Then each encoding region $r_i$ would be infilled with a set of parallel line segments of the encoded angle $\delta_i$ and the interval of linewidth $w$, whose endpoints are on the boundary of the encoding region. Calculating coordinates of these endpoints, we obtain \texttt{XY} values of instructions guiding the 3D printer to fabricate infilling lines. The value of \texttt{Z} is usually set as a little higher than the height of one 3D printed layer and kept constant, for example, $0.3mm$ when the layer height is $0.2mm$. After determining the infilling path \texttt{XYZ}, the corresponding extrusion amount \texttt{E} should ensure the 3D printer fabricates the target SCS microstructure along infilling paths. Specifically, the extrusion amount should be set to fabricate the infilling line as a half-cylinder, rather than a rectangle in the common 3D printer setting. Considering the 3D printing process in Figure~\ref{gcode_cal}, the filament with a diameter of $d_f$ is fed into the 3D printer with the nozzle of diameter $d_n$. And then the 3D printer prints infilling lines with a height of $h$ and linewidth of $w$. Because the amount of filament fed and extruding should be equal, the relationship between the fed filament length $l_f$ and the infilling line length $l_i$ is as follows:
\begin{equation}
    \begin{aligned}
    \pi(d_f/2)^2 l_f &= \pi(hw/2)/2 l_i \\
    l_f&=hw/{d_f}^2\ \cdot\ l_i,
    \end{aligned}
\end{equation}
where linewidth $w$ is set the same as the nozzle diameter $d_n$ and the infilling line length $l_i$ is the distance between the coordinates of the current instruction and the previous one. Calculating $l_i$ and then we obtain $l_f$. Considering the mobility of the molten material and hardware differences between different 3D printers, $l_i$ should be multiplied by one factor to generate the final value of $E$. The factor is usually less than $1$ to avoid squeezing and maintain the surface tension of molten material to generate the SCS microstructure. Finally, the set of instructions is saved to one \texttt{G-code} file, which can guide the 3D printer to fabricate the corresponding AnisoTag.

%% file: sections/detection.tex
\section{Detection Prototype}
\label{detection}
The AnisoTag detection prototype is the implementation of the theoretical model shown in Figure~\ref{coordinate_system}(a). As mentioned in Section~\ref{anisotropy_analysis}, the data is encoded into SCS microstructures with different angles on AnisoTag, which reflects different illumination patterns under the laser beam. Thus, if we can determine which the reflected illumination pattern is, we can infer the embedded information on the AnisoTag based on the mapping relationship. 
In this section, we first analyze the reflection model, derive a clear relationship between the reflected illumination patterns and the SCS microstructures, and infer the characteristics of these patterns. Based on the concluded relationships and characteristics, we design an easy-to-implement method to detect reflected illumination patterns. 
Then, we illustrate the implementation details of the detection prototype and the corresponding detection algorithm.

\subsection{Reflected Illumination Pattern Analysis}
\label{anisotropy_analysis_2}
Based on the model we proposed in Section~\ref{anisotropy_analysis}, we summarize some features of the reflected illumination patterns.

First, we observe that these reflected illumination patterns pass through the same fixed point, regardless of how their shapes change with the direction of the SCS microstructure. The reason is that the top surface of the SCS microstructure is always parallel to the $XoY$ plane, resulting in the light reflected on that top surface illuminating at the same point. In that case, the angle between the normal of the reflecting surface and the $Z$-axis is $\theta = 0$. Substituting $\theta = 0$ into Eq.~(\ref{reflected_vector}), the vector of the reflected ray is $\vec{b}=(0,\sin\alpha,\cos\alpha)$, which is only related to the incident angle $\alpha$. And according to the position of the background plane $P:y=d$, the coordinates of the fixed point could be calculated as $(0, d, d/\tan\alpha)$.

Second, we notice that the reflected illumination pattern is part of a conic section, whose parameters vary with the direction of the SCS microstructure. Based on Eq.~(\ref{reflected_vector}), the angle between the reflected ray $\vec{b}$ and the direction of cylindrical axis $\vec{c}$ could be calculated as $\arccos (\cos\phi\sin\alpha)$; see Appendix for derivation process. Thus, for a given incident angle $\alpha$ and the cylindrical section angle $\phi$, the reflected rays form a light circular cone with origin $O$ as its vertex and the cylindrical axis $\vec{c}$ of the SCS microstructure as its axis. Then, the reflected illumination pattern $\mathcal{I}$ is the intersection of that light cone with the background plane $P: y=d$, so $\mathcal{I}$ is a conic section. After determining the functional form of the reflected illumination patterns $\mathcal{I}$, using the relevant theorem, we can derive its function about parameter settings $\mathcal{I}=f(\alpha, \phi, d)$; see Appendix for the derivation process. Thus, based on the function between the reflected illumination pattern $\mathcal{I}$ and the encoded angle $\phi$ of SCS microstructure on AnisoTag, we could detect these patterns and then decode the embedded data on AnisoTag.

Combining the above two conclusions we found that as the direction of the SCS microstructure changes, the reflected illumination pattern $\mathcal{I}$ could be regarded as a curve line segment rotating around the fixed point. Moreover, this inference is still applicable when the microstructure is not strictly cylindrical but a convex surface, because the convex microstructure has the same flat top surface and is equivalent to a partially cylindrical microstructure on the macroscopic scale of the detection prototype. Drawing a circle around the fixed point, the reflected illumination pattern intersects with that circle at two points, and these points are different for different reflected illumination patterns. Thus, the reflected illumination pattern could be uniquely determined by detecting two intersection points. Compared with determining the pattern by sampling a large number of points, this inference significantly reduces the complexity and hardware requirements of the detection process. 

\begin{figure*}
\setlength{\belowcaptionskip}{0pt}
\setlength{\abovecaptionskip}{3pt}
\begin{center}
  \includegraphics[width=\linewidth]{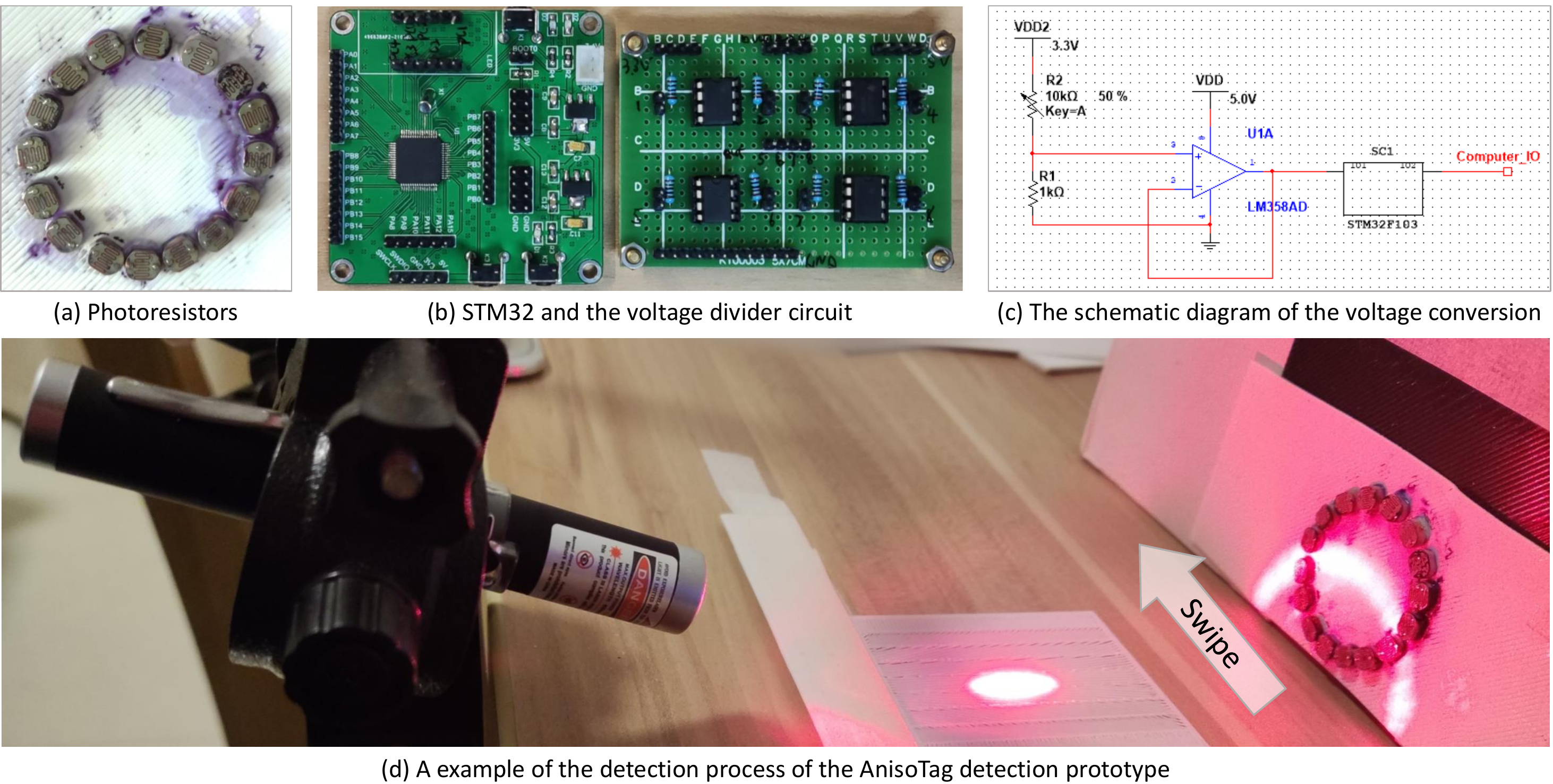}
  \caption{\textbf{Implementation details of AnisoTag detection prototype.} The photoresistors in (a) would be connected to the voltage divider circuit on the right side of (b) and its voltage is digitized and transmitted to the computer by the STM32 on the left side of (b). In (c), $R2$ represents the photoresistor. In the detection process (d), hardware circuits are put into a box for the convenience of the experiment.}
  \Description{
  Implementation details of AnisoTag detection prototype. The photoresistors in (a) form a circle. (b) shows two circuit boards. In (d), the laser pointer pen is held by one tripod, lighting on the AnisoTag on the table and reflected as the illumination pattern of the curve line shape on the photoresistor array of the detection prototype.
  }
  \label{fig:prototype}
\end{center}
\end{figure*}

\subsection{Detection Prototype Implementation}
\label{sec:detection_prototype}
Based on the design in the previous subsection, the detection prototype of AnisoTag is implemented as Figure~\ref{fig:prototype}. We use photoresistors as the optical sensors and place $16$ photoresistors evenly along a circle on the 3D printed background plane; see Figure~\ref{fig:prototype}(a). Then, we construct a signal transmission module based on STM32, a 32-bit Flash microcontroller family developed by ST Microelectronics, to convert and transmit the photoresistor voltage to the computer; see Figure~\ref{fig:prototype}(b).\footnote{We use two voltage divider circuits in the detection prototype.} Specifically, each photoresistor is connected to a $1k\Omega$ resistor, and they divide the total voltage of $3.3V$. After passing through a voltage follower, the divided voltage of the photoresistor is analog-to-digital (AD) converted by the STM32 to digital voltage data of $12$ bits and then transmitted to the computer. A schematic of this process is shown in Figure~\ref{fig:prototype}(c). Since one AD conversion takes $25\mu s$, voltage values of $16$ photoresistors in sequence take a total of $0.4ms$. According to the product description of STM32, the highest rate of error-free transmission rate between STM32 and the computer is $1.152Mbps$, which means that a set of voltage data can be transmitted within $0.25ms$. 

Then, we use a laser pointer pen as the incident source and the AnisoTag reflects illumination patterns on the background plane; see Figure~\ref{fig:prototype}(d). Before the detection process, we record some sets of voltage data detected by the prototype, including the ambient voltage data $V_A$ in the current light condition and the voltage data $V_i$ irradiated by the reflected illumination pattern $\mathcal{I}_i$. Each voltage data has $12\times 16$ bits. Then in the detection process, for the received voltage data $V^\prime$, the similarity $S_i$ between $V^\prime$ and the voltage data of the reflected illumination pattern $\mathcal{I}_i$ could be calculated as:
\begin{equation}
    S_i = corr(V^\prime-V_A, V_i-V_A),
\end{equation}
where $corr(\cdot)$ calculates the correlation of two inputs. Then, $V^\prime$ would be detected as the reflected illumination pattern of state $i$:
\begin{equation}
    i = \arg \max \ S_i, \quad  if\quad max(S_i) > T,
\end{equation}
where $T$ is a detection threshold, which is usually set to $0.9$ in the implementation. In the detection process, AnisoTag is swiped from one side of the laser-illuminated area to the other. Between each of the two encoding regions of AnisoTag, the 3D print trajectory leads to the accumulation of 3D printing material here, forming a borderline slightly higher than the SCS microstructure; see Figure~\ref{gui_workflow}. Different from the clear curve illumination pattern reflected by encoding regions, the pattern reflected by these borderline is diffuse reflection, whose maximum similarity $max(S_i)$ would be smaller than threshold $T$ and would be considered as an invalid detection result. Thus, these lower photoelectric signal responses can be adopted to separate adjacent encoding regions. On the other hand, the detection result would be recorded only if the maximum similarity $max(S_i)$ is greater than $T$, i.e., the AnisoTag reflects a clear enough reflected illumination pattern. Based on the mapping relationship between the reflected illumination pattern and the encoded data, we can extract the bitstream embedded in AnisoTag. One example video is provided in the supplementary material.

%% file: sections/experiments.tex
\section{Experiment Evaluation}
\subsection{Experimental Methodology}
\label{implementation_details}
In the AnisoTag fabrication process and detection process, many parameters are adjustable. We present default parameter settings in this subsection and modified part of them for specific evaluation in the following subsections.

The default parameter settings are described below. For experimental and illustrative simplicity, we fabricate AnisoTag individually with an easily accessible size setting: $53.98\times 85.6 mm$, the same as a credit card. Without loss of generality, we divide $n=17$ encoding regions along the horizontal axis, where the encoding region has a size of approximately $85.6/17\approx5mm$ and could encode $m=3$ bits. Then, using the AnisoTag \texttt{G-code} tool with a bitstream of $17\times3=51$ bits and above parameters as input, we obtain a \texttt{G-code} file, which is fed into an FDM 3D printer of brand Creality Ender-3 V2 (\$262) to fabricate AnisoTag with white PLA filament (\$20/kg) at a recommended 3D printing speed $50 mm/s$. For AnisoTag with a single layer, each AnisoTag takes about $1g$ of PLA materials, and about $15\%$ more 3D printing time compared to plain 3D printed tags because of the complex extruding trajectory. The detection process of AnisoTag is implemented in the indoor environment with an ambient illumination intensity of around $400 lux$. The detection prototype is set as Figure~\ref{fig:prototype}(d). In the detection prototype, we use an STM32F103 ($\$7$) as the microcontroller. And the collimated laser source of the detection prototype is the most common Class IIIA laser pointer pen ($\$3$) with only a small potential hazard, whose frontal illuminated spot is $5 mm$ in diameter. We fixed the position of the laser source using a tripod and set the incidence angle to $70^\circ$. In the background plane, $16$ photoresistors are uniformly placed around a circle with a radius of $1.5cm$. The distance between the background plane and the point of incidence is set to $d=6.5cm$, where the reflected fixed point coincides with the center of the circle surrounded by photoresistors. Swiping the AnisoTag from one side of the laser-illuminated area to another side, we record the detected bitstream on the computer and calculate its bit error rate.

\subsection{Visual Quality}
Before evaluating the performance of AnisoTag, we compare its visual quality with similar methods, including LayerCode \cite{layercode} and acoustic barcodes \cite{acoustic_barcodes}, which could be fabricated with consumer-grade 3D printers. All these tags are fabricated with the same size, same 3D printer, and same material as the AnisoTag default settings. Yet they have different information capacities due to different data encoding methods. Specifically, on the same-size tag, AnisoTag encodes $51$ bits, LayerCode encodes $25$ bits, and acoustic barcode encodes $40$ bits. We evaluate the visual quality in the case of AnisoTag embedded with more information.

\begin{figure}[t]
\setlength{\belowcaptionskip}{0pt}
\setlength{\abovecaptionskip}{0pt}
    \begin{center}
    \subfloat[LayerCode]{
        {\centering\includegraphics[width=1 in]{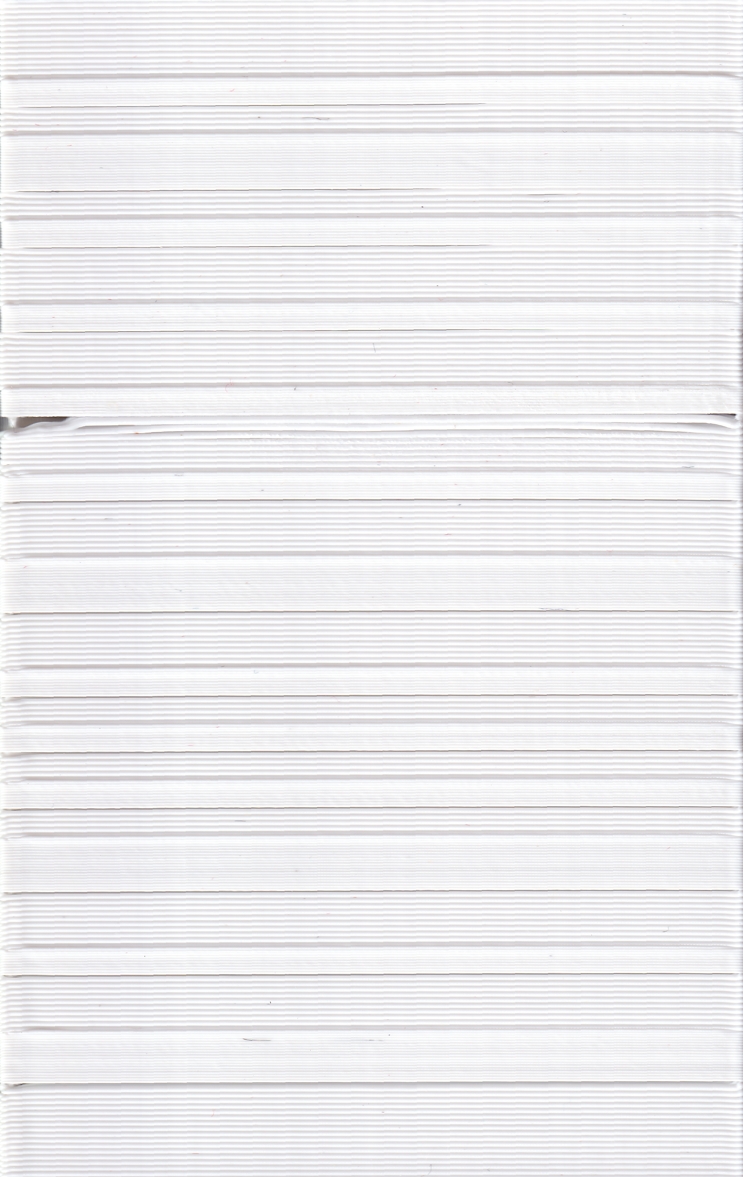}
        }
    }
    \subfloat[Acoustic barcode]{
        {\centering\includegraphics[width=1 in]{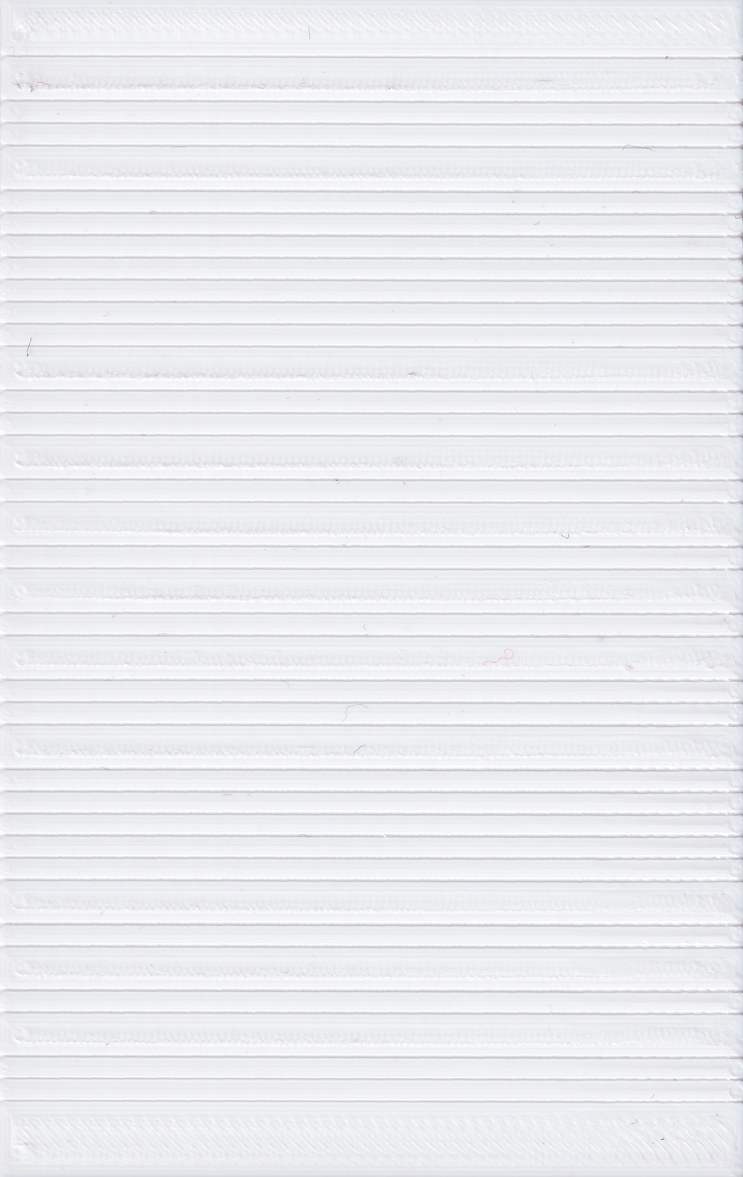}
        }
    }
    \subfloat[AnisoTag]{
        {\centering\includegraphics[width=1 in]{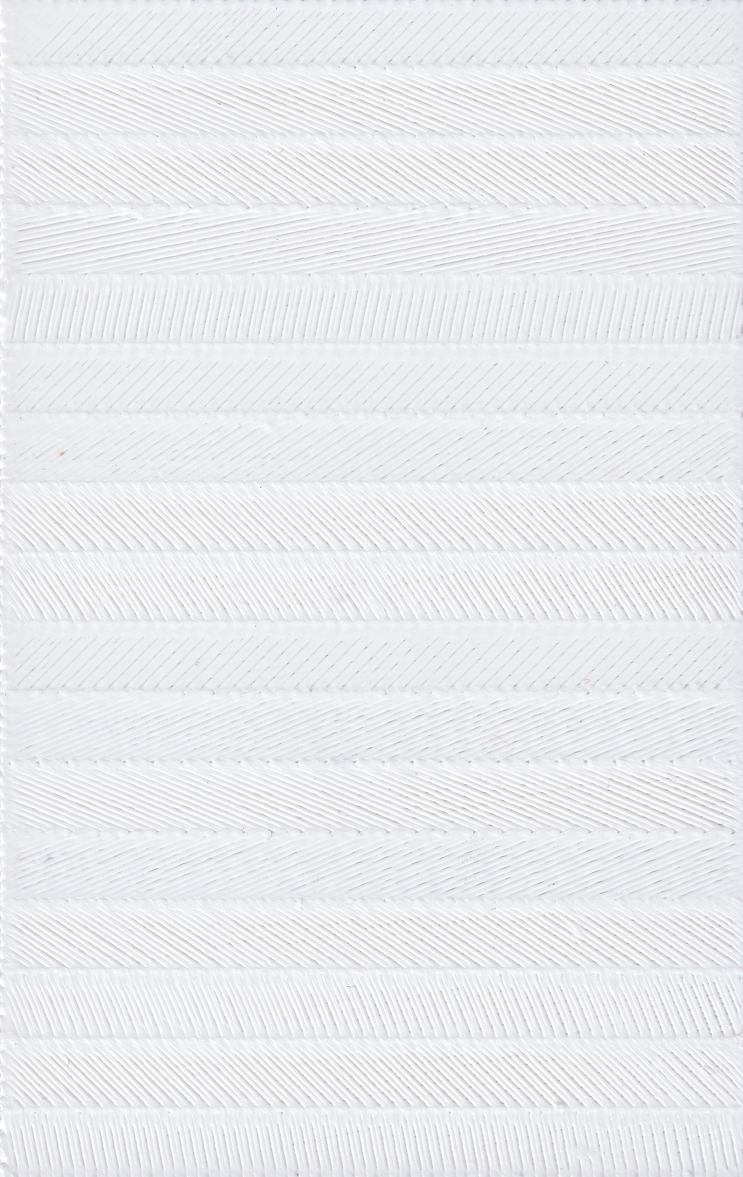}
        }
    }
    \end{center}
\caption{\textbf{Visualization of some 3D printed tags fabricated by available methods.} Compared with (a) and (b), artifacts in (c) AnisoTag are more unobtrusive.}
\Description{The artifacts on AnisoTag are more unobtrusive compared to LayerCode and Acoustic barcode.}
\label{visual_comparision}
\end{figure}

First, for each method, we fabricate the corresponding tag $3$ times with random data to obtain average metrics in the subsequent evaluations. Then, we use a paper scanner to obtain their surface appearance to reduce the environmental impact, such as shadows caused by illumination. To simulate human eye perception, we introduce the field of view (FOV) to calculate the number of pixels these tags should have at a comfortable viewing distance. Considering that the angular resolution of the human eye is typically around $1 arcminute$, at the comfortable viewing distance of $25cm$, a $53.98\times85.6mm$ tag is equivalent to a $706\times 1118$ resolution image to human eyes. Thus, the scanned images of 3D printed tag are cropped and resized to the above size; see Figure~\ref{visual_comparision}. Then, we have two metrics to evaluate their visual quality. One is the visual saliency-induced index (VSI) \cite{VSI}. As shown in Table~\ref{tab:visual_quality}, we calculate the VSI between them and one pure white image, thus the 3D printed tag with a \emph{higher} VSI is closer to one ideal pure white tag. Another metric is edge percent, which would be \emph{lower} if the tag has fewer perceptible patterns on its surface. To implement the edge detection, we use the Sketch Tokens \cite{ST} to obtain the edge possibility of the scanned tag image and calculate the percent whose possibility is larger than $0.2$ as the edge percent. As shown in Table~\ref{tab:visual_quality}, AnisoTag has a relatively better visual quality with fewer bumps on its surface.

\begin{table}[h]
\caption{Visual quality comparison between AnisoTag and similar methods.}
\Description{This table has 3 rows and 4 columns. The first row lists the method names, including LayerCode, Acoustic barcode, and AnisoTag from left to right. The second row illustrates the VSI value: 0.603, 0.671, and 0.714 from left to right. The last row lists the edge percent: 3.124, 3.6, and 0.151 from left to right.}
\label{tab:visual_quality}
\begin{center}
\resizebox{\linewidth}{!}{
\begin{tabular}{lcccc}
\toprule
Methods       & LayerCode & Acoustic barcode & AnisoTag \\
\midrule
VSI $\uparrow$    & 0.603 & 0.671 & \textbf{0.714} \\
\midrule
Edge $\downarrow$ (\%)     & 3.124 & 3.600 & \textbf{0.151} \\
\bottomrule
\end{tabular}
}
\end{center}
\end{table}

\subsection{The Fabrication Settings}
Among the many parameters used in the AnisoTag fabrication process, there are two main parameters that affect the accuracy of data extraction: the number of encoding regions $n$ and the number of encoding bits $m$ per region. Both of them determine the information capacity of AnisoTag.

\subsubsection{The impact of encoding regions number}
\begin{figure}[t]
  \begin{center}
  \includegraphics[width=3.3in]{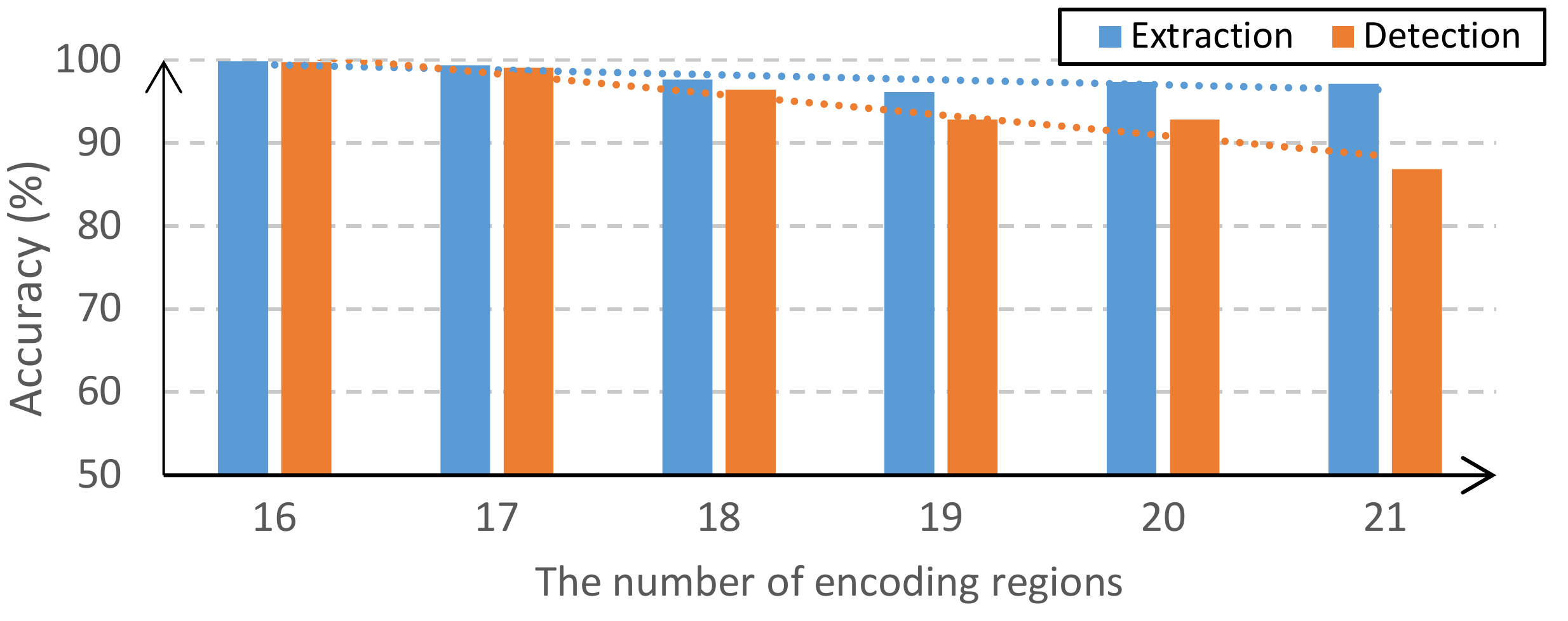}
  \caption{The impact of the number of AnisoTag encoding regions on the data detection and extraction accuracy.}
  \Description{The line chart is plotted with the number of encoding regions as the x-axis and the detection and extraction accuracy as the y-axis. Along with the encoding regions increasing from 16 to 21, the extraction accuracy declines a little but is still higher than 98\%. The detection accuracy declines are close to 100\% when the number of encoding regions is 16 and 17. Yet when encoding regions are more than 17, detection accuracy declines and only has 86\% when the number of encoding regions is 21.}
  \label{fig:encoding_region}
  \end{center}
\end{figure}

Keeping all other settings the same, we increase the number of encoding regions from $16$ to $21$. As each region encodes $3$ bits, the information capacity of the testing AnisoTag ranges from $48$ to $63$ bits. For each setting, we fabricate $5$ AnisoTags embedded with a random bitstream of the corresponding length. And for each AnisoTag, we execute the detection process $5$ times. The detection results of $25$ experiments are used to calculate the average accuracy of AnisoTag under each setting. As shown in Figure~\ref{fig:encoding_region}, we demonstrate two types of accuracy: data extraction accuracy and data detection accuracy. Detection success means that the decoded data from AnisoTag conforms to the pre-defined data format and that success rate is the detection accuracy. Extraction accuracy is the bit recovery rate in the case of successful detection. The accuracy, especially the detection accuracy, decreases as the number of encoding regions increases and drops rapidly when the number of encoding regions exceeds $17$. As mentioned in Section~\ref{implementation_details}, when the encoding region number is set to $17$, its width is approximately $5mm$, which is roughly the same as the diameter of the laser beam used in the detection prototype. When we encode AnisoTag with more than $17$ encoding regions, the encoding region would be narrower than $5mm$. As a result, the laser beam with a diameter of $5mm$ is reflected not only by the individual encoding region but also by the neighboring areas, thus creating noise signals in the reflected illumination pattern. These noise signals reduce the similarity response of the reflected illumination pattern during the detection process, which interferes with the bitstream synchronization of the detection procedure. But the above phenomenon has little effect on the extraction process, whose accuracy maintains over $96\%$ for all numbers of encoding regions. To keep both accuracy high, we use $17$ encoding regions as the default AnisoTag setting.

\subsubsection{The impact of encoded bits number}
\begin{figure}[t]
  \begin{center}
  \includegraphics[width=3.1in]{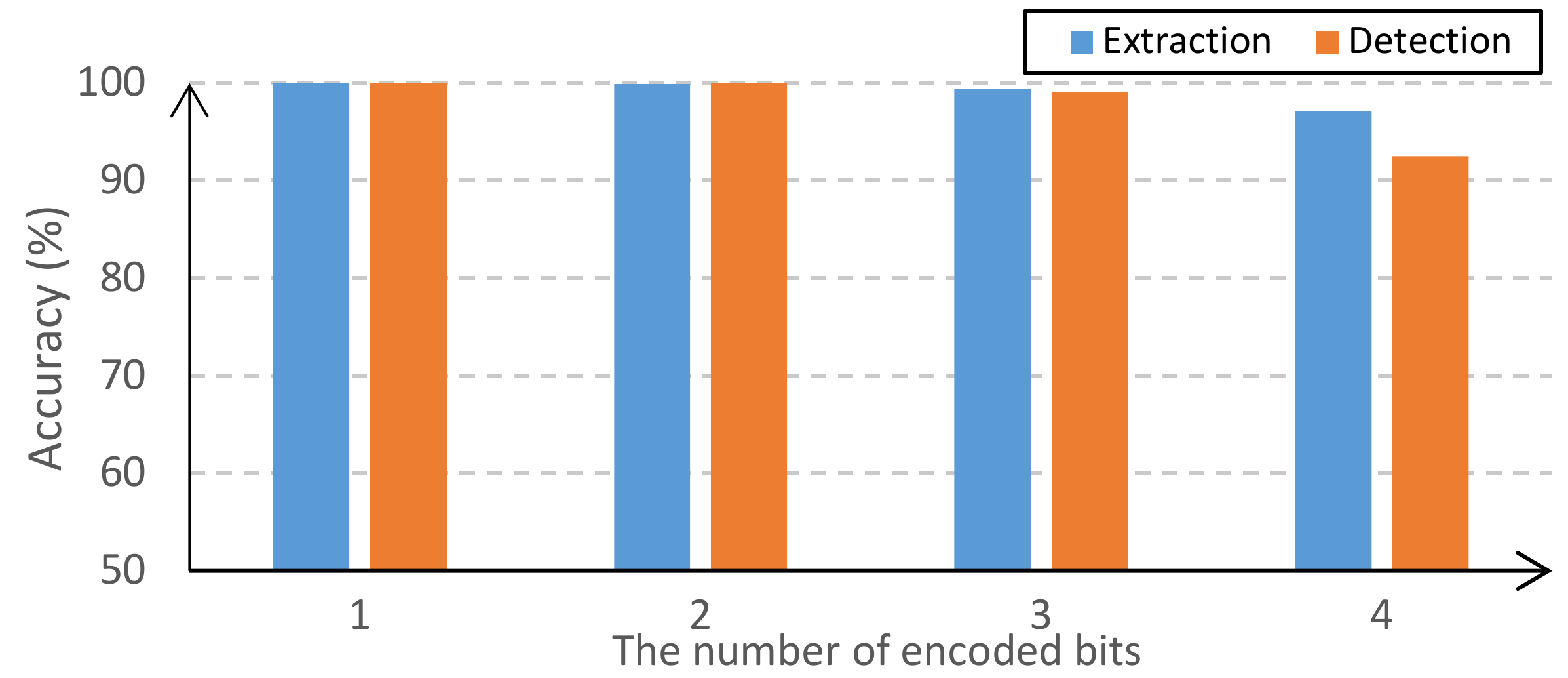}
  \caption{The impact of the number of encoded bits per encoding region of AnisoTag on the data detection and extraction accuracy.}
  \Description{The line chart is plotted with the number of encoded bits per encoding region as the x-axis and the detection and extraction accuracy as the y-axis. Along with the encoding bits increasing from 1 to 3, the extraction accuracy and detection accuracy are affected a little and higher than 99\%. Yet when the number of encoding bits is 4, the detection accuracy is 92\% and the extraction accuracy is 96\%.}
  \label{fig:encoded_bits}
  \end{center}
\end{figure}

Similarly, we evaluate the effect of the number of encoded bits per encoding region on AnisoTag. The number of encoded bits varies from $1$ to $4$, corresponding to $2$, $4$, $8$, and $16$ types of reflected illumination patterns and $17$, $34$, $51$, and $68$ bits of information capacities. The experimental results are shown in Figure~\ref{fig:encoded_bits}. AnisoTag with fewer encoded bits would reflect easily distinguished illumination patterns, resulting in data detection and extraction accuracy close to $100\%$. However, when encoding $4$ bits on each region, the detection performance degrades. We believe the reason is that the limited $16$ photoresistors used in the AnisoTag detection prototype cannot fully sample $16$ types of reflected illumination patterns. Therefore, by default, we encode $17\times 3=51$ bits of data on one AnisoTag. In addition, we note that the detection performance is limited by the detection prototype, which means that the detection accuracy and information capacity of the AnisoTag could be further improved with better hardware, such as a finer laser source or more photoresistors.

\subsection{Gray-coding}
As mentioned in Section~\ref{encoding}, we use Gray code \cite{gray_code_wiki} to map the bitstream and the angle of SCS microstructure such that the adjacent angles correspond to bitstreams with only one-bit difference. Table~\ref{tab:gray_coding} takes the detection results of the previous subsection to illustrate the effectiveness of Gray coding. Without using Gray code, the bit error rates (BER) becomes much higher, for a common detection error in the AnisoTag detection prototype is to mistake one reflected illumination pattern for another adjacent one, while the common binary encoding may lead to several bit errors. For example, mistaking the reflected illumination pattern of type $3$ to type $4$ would cause the decoded bitstream to change from $011$ to $100$ under binary coding. But under the Gray coding, the same mistake only changes bitstream from $010$ to $110$.

\begin{table}[h]
\caption{The BER (\%) of AnisoTag w/ and w/o Gray coding when encoding different numbers of bits in one encoding region.}
\Description{When the number of encoding bits is 1, 2, 3, and 4, the BER in extraction with Gray coding is 0, 0.078, 0.627, and 2.902. The BER without Gray coding is 0, 0.157, 1.255, and 7.373.}
\label{tab:gray_coding}
\begin{center}
\resizebox{\linewidth}{!}{
\begin{tabular}{lcccc}
\toprule
Number of encoded bits  & 1 & 2      & 3     & 4     \\
\midrule
With Gray coding    & \textbf{0} & \textbf{0.078} & \textbf{0.627} & \textbf{2.902} \\
W/o Gray coding & \textbf{0} & 0.157  & 1.255 & 7.373 \\
\bottomrule
\end{tabular}
}
\end{center}
\end{table}

\begin{table}[h]
\caption{Comparison of extraction ability of AnisoTag fabricated with types of 3D printers.}
\Description{The BER of AnisoTag fabricated by 3D printer with brands of JG, Creality, and Ultimaker is 1.176, 0.863, and 0.588.}
\label{tab:brands}
\begin{center}
\resizebox{\linewidth}{!}{
\setlength{\tabcolsep}{.08in}{
\begin{tabular}{lccc}
\toprule
Brand    & JG     & Creality & Ultimaker \\
\midrule
Bit error rate (\%) & 1.176 & 0.863 & 0.588 \\ 
\bottomrule
\end{tabular}
}
}
\end{center}
\end{table}

\subsection{3D Printers}
To evaluate the impact of 3D printers on AnisoTag, we fabricate AnisoTags with default settings using $3$ brands of FDM 3D printers: JG Maker Z-603S, Creality Ender-3 V2, and Ultimaker S3. The first two are single extrusion 3D printers using filaments of diameter $1.75mm$. The last one uses $2.85mm$ diameter filaments and has dual extruders. Yet we only use one extruder to fabricate AnisoTag in experiments. Similarly, we fabricate $10$ AnisoTags with each 3D printer and extract their data $5$ times. 
The experimental results shown in Table~\ref{tab:brands} demonstrate that the proposed AnisoTag could be fabricated on different brands of 3D printers and maintain good performance with BER below $2\%$.

\subsection{3D Printing Materials}
A wide variety of 3D printing materials can be used to fabricate AnisoTags. To confirm this, we fabricate AnisoTags with various colors and types of materials and calculate the extraction results of embedded data.

\begin{figure}[t]
  \begin{center}
  \includegraphics[width=3.3in]{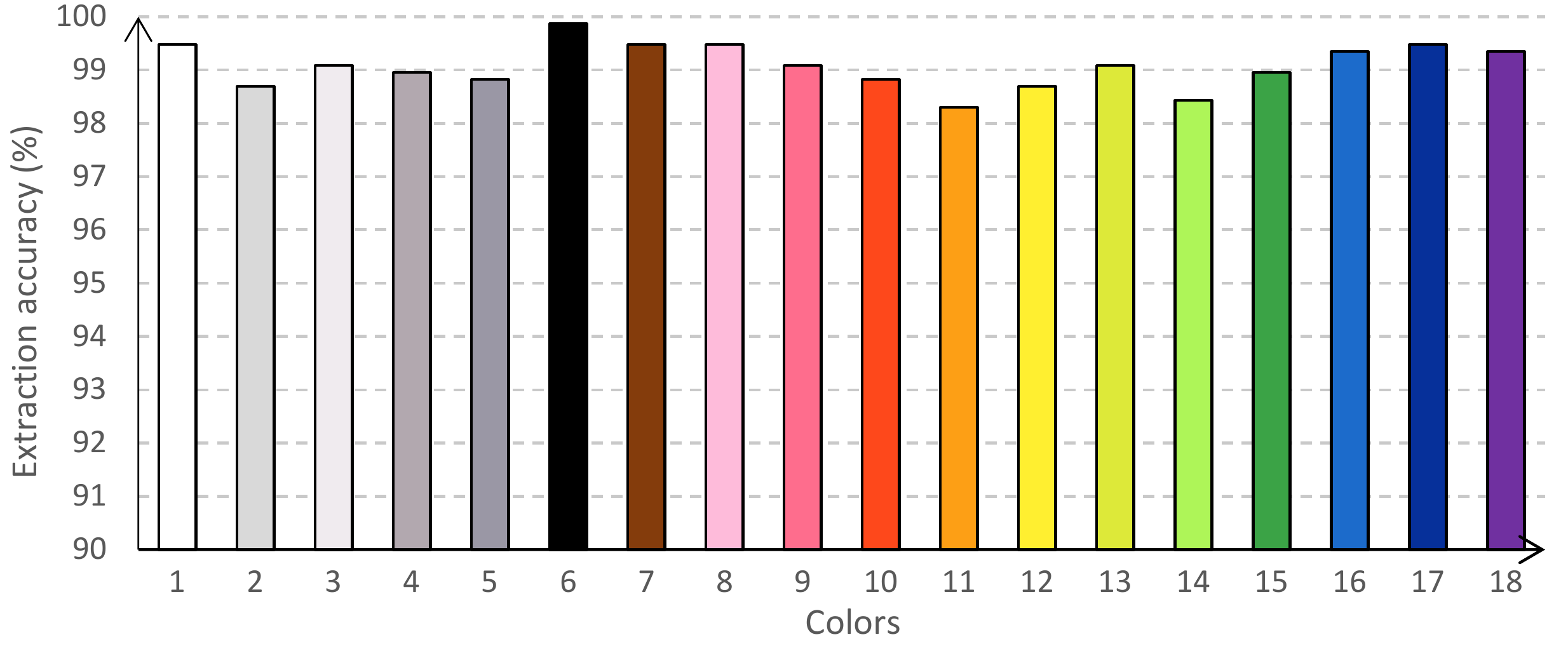}
  \caption{The impact of color of materials on extraction accuracy.
  }
  \Description{
  The AnisoTag fabricated by materials of all testing colors has an extraction accuracy of no less than 98\%.
  }
  \label{fig:acc_color}
  \end{center}
\end{figure}

\subsubsection{The impact of materials colors}
We use $18$ colored materials to fabricate AnisoTag. All materials are colored PLA except the transparent and the semi-transparent AnisoTags fabricated by PETG (label $2$ and $3$ in Figure~\ref{fig:acc_color}). We fabricate $3$ AnisoTags of each color and extract each AnisoTag $5$ times to obtain the average extraction accuracy; see Figure~\ref{fig:acc_color}. The experimental results show that the data embedded in AnisoTags of all colors could be extracted with accuracy over $98\%$, proving the superior usability of AnisoTag. Moreover, two other points are worth discussing. First, we find that black materials are more suitable for fabricating AnisoTag. Compared to the white one, black AnisoTag has less diffuse reflection, resulting in clearer reflected illumination patterns and higher extraction accuracy; see Figure~\ref{fig:acc_color}. Second, transparent materials could also be used in AnisoTag fabrication, whose transparency does not have much influence on the detection process of AnisoTag. When the laser beam irradiates on transparent AnisoTag, a portion of the laser beam is reflected as an illumination pattern, and another passes through it and is then diffusely reflected by the tabletop underneath the AnisoTag. The interference of that diffuse reflection is tolerable for detecting prototypes.

\begin{figure}[t]
  \begin{center}
  \includegraphics[width=3.2in]{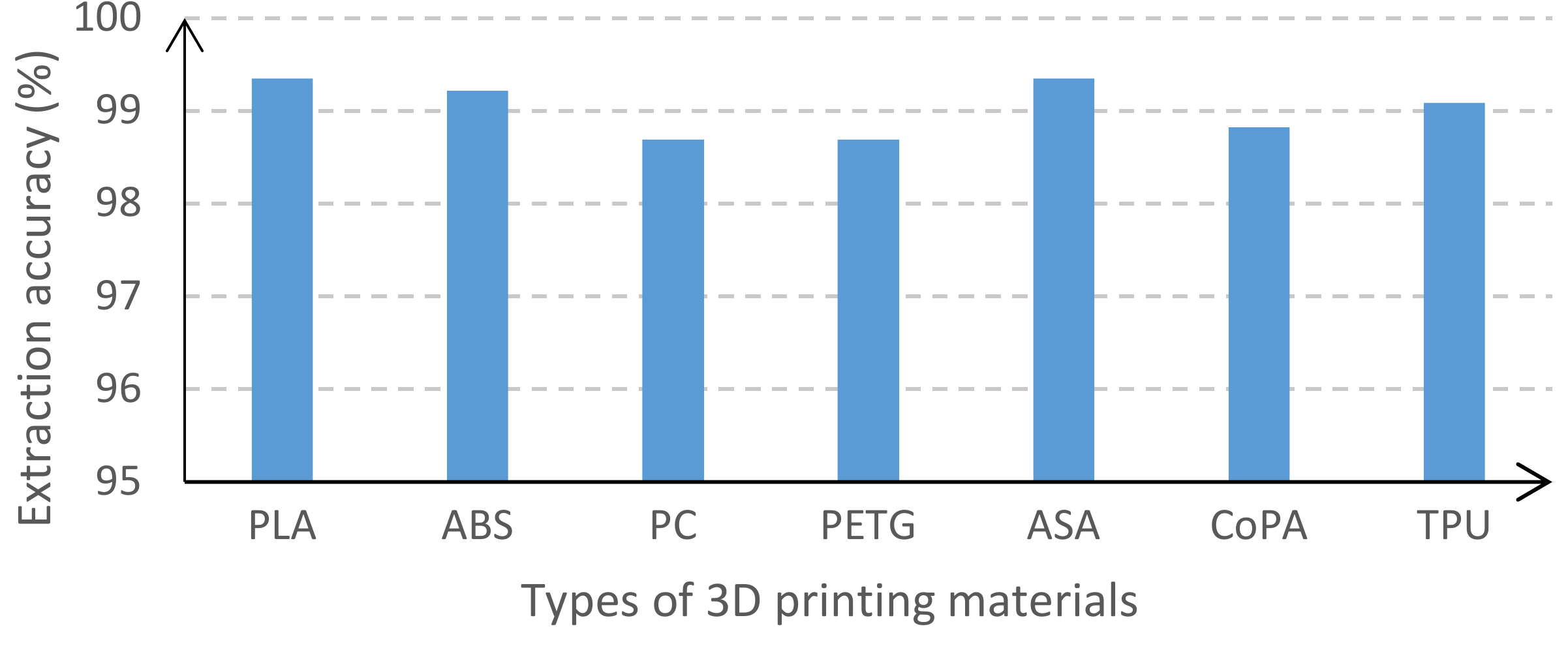}
  \caption{The impact of types of materials on extraction accuracy.
  }
  \Description{
  The AnisoTag fabricated by all testing materials has an extraction accuracy of no less than 98.5\%.
  }
  \label{fig:types}
  \end{center}
\end{figure}

\subsubsection{The impact of materials types}
There is a wide type variety of 3D printing materials for users to choose from based on their demands of durability, flexibility, price, etc. To evaluate the generalizability of AnisoTag in terms of 3D printing material types, we fabricate them with PLA, ABS, PC, PETG, ASA, CoPA Nylon, and TPU filaments. The colors of these materials are not the same, including black, white, and orange, but the experimental results in the previous subsection demonstrate that the color of the filament has little effect on the extraction accuracy of AnisoTag. For each material, we fabricate $3$ AnisoTags and extract them $5$ times to obtain the average extraction accuracy. As shown in Figure~\ref{fig:types}, the extraction accuracy of AnisoTags fabricated by the above materials exceeds $98.5\%$. In summary, AnisoTag can be fabricated with a variety of types and colors of materials.

\section{Extended Applications}
\label{sec:extend}
In the AnisoTag fabrication, we use our \texttt{G-code} tool rather than common 3D slicing software to generate the 3D printing file, thus we could implement multiple 3D print setups on a single 3D printed layer and fabricate objects with reflection anisotropy. Besides the AnisoTag system, this technique could benefit other applications.

\subsection{AnisoBarcode}
Considering that 3D printed products like figurines commonly have a bottom plane, it is possible to extend our planar-based data embedding method to the bottom plane of 3D printed objects with necessary modifications. For example, we can reverse the 3D printing orientation with the bottom side up and then fabricate SCS microstructures. The other alternative is leaving the space in the bottom 3D printing layer and fabricating the AnisoTag pattern in the higher layer of the corresponding position. Then, SCS microstructures would not directly contact the print platform for smooth cylindrical surfaces, and the rest of the bottom layer sticks to the platform for well 3D printing. 

\begin{figure}[t]
  \begin{center}
  \includegraphics[width=\linewidth]{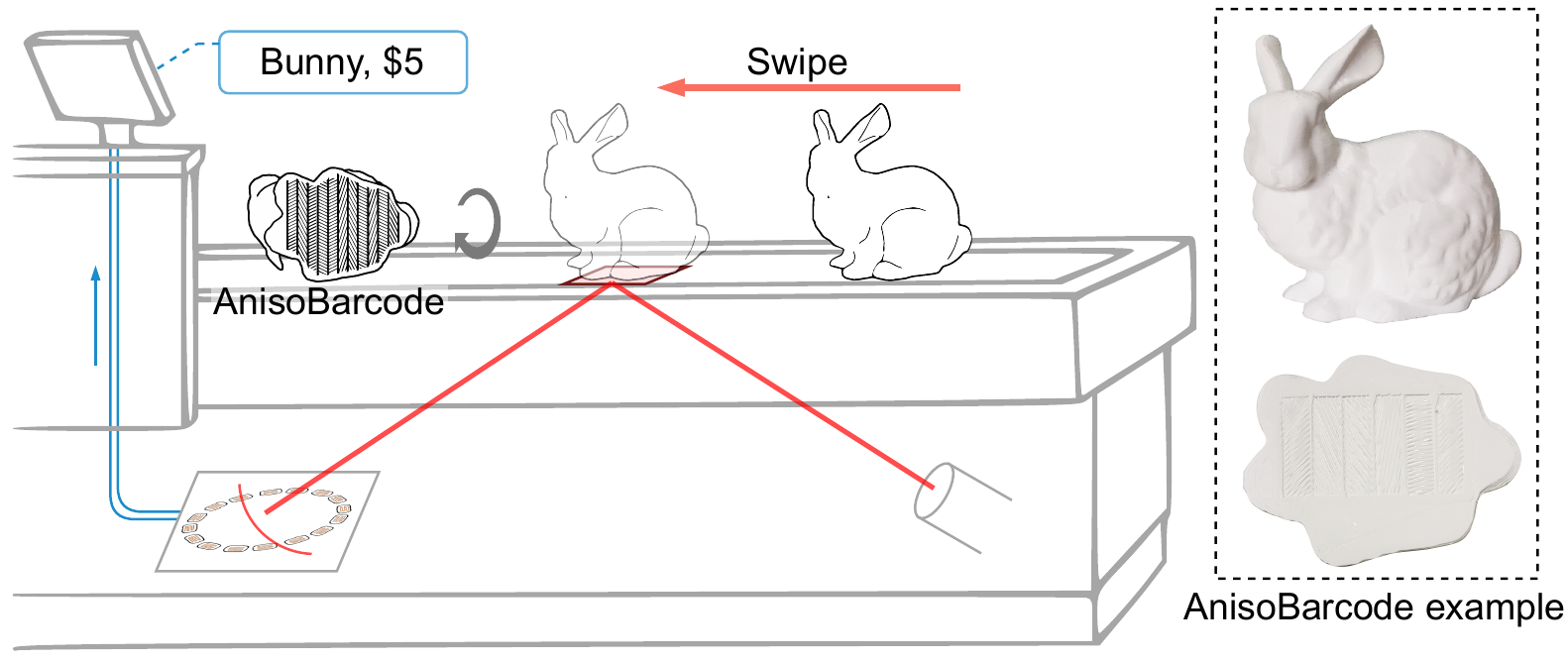}
  \caption{\textbf{The schematic of the application scenario for AnisoBarcode.} The AnisoBarcode is fabricated on the bottom surface of the 3D printed products and the clerk could quickly obtain the product information by swiping it over the detection prototype inside the counter. An example of 3D printed AnisoBarcode on the bottom surface is shown on the right.
  }
  \Description{
  The picture shows a cashier counter where 3D printed goods with AnisoBarcode on the bottom will slide from the right side to the left side, and a detection prototype inside the counter will shine a laser through the transparent top of the counter to the bottom of the goods and reflect it to the light sensor to get the information of the goods.
  }
  \label{fig:Anisobarcode}
  \end{center}
\end{figure}

As a result, SCS microstructures could be fabricated on the bottom surface of most 3D printed objects for metadata embedding or watermarking. One of the interesting scenarios is in the trade of 3D printed artwork, where we can use AnisoTag techniques to design AnisoBarcode as merchandise barcodes for fast item identification; see Figure~\ref{fig:Anisobarcode}. The detection prototype is set under the counter and the cash register can quickly obtain the product information when the 3D printed artwork with AnisoBarcode swipes over the counter. Compared to other 3D printed tagging methods, AnisoBarcode has low-complexity fabrication and real-time extraction, which meet the needs of this scenario. For embedded chips or tagged barcodes, AnisoBarcode could be fabricated along with the object in the 3D printing process, which reduces the fabrication complexity and avoids the failure of paper tags due to abrasion.

\subsection{High-Capacity Tag}
\label{sec:highcapacitytag}
\begin{figure}[t]
  \begin{center}
  \includegraphics[width=3.3in]{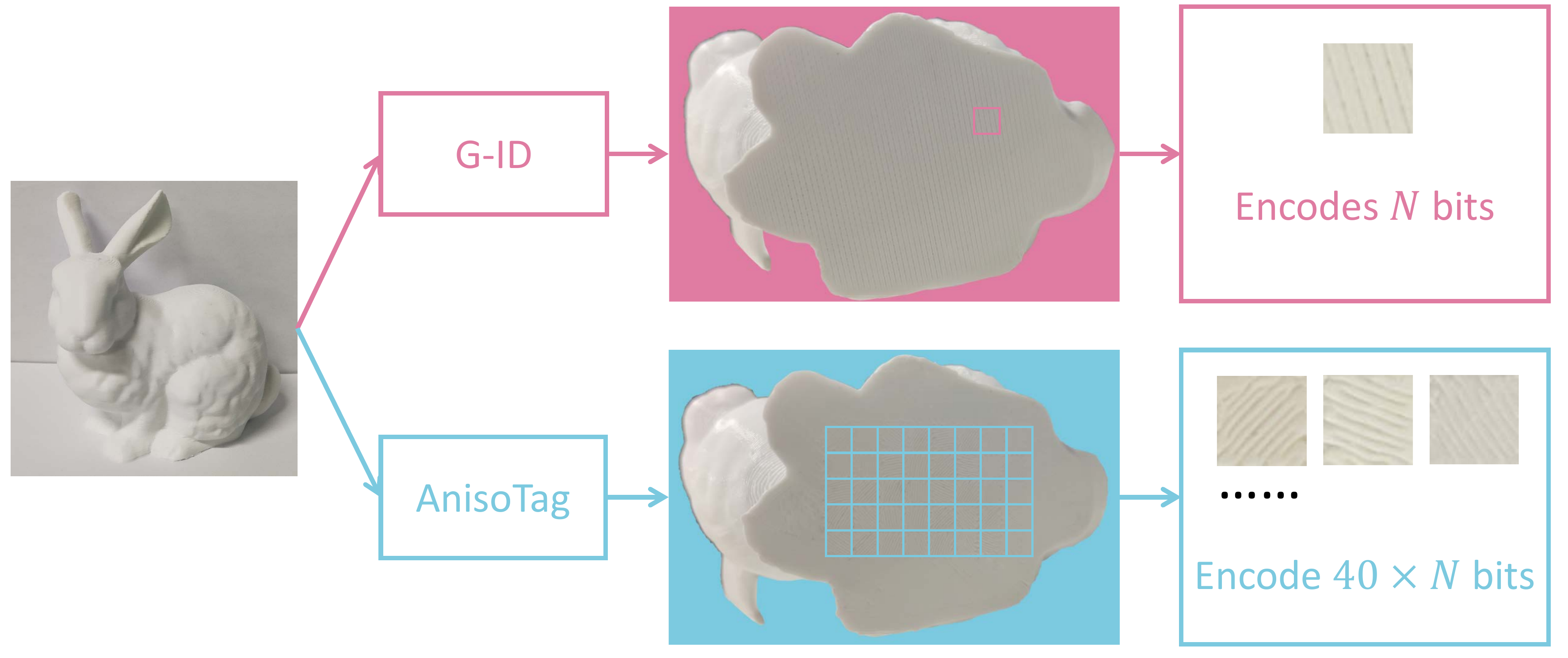}
  \caption{\textbf{An example of high-capacity 3D printed tag.} With our \texttt{G-code} tool, tags with multiple infilling angles could be attached to the 3D printed object and could be extracted with image processing operations similar to existing 3D printed tagging methods.
  }
  \Description{
  On the  bottom layer of 3D printed objects, G-ID uses only one infilling angle but AnisoTag could use up to 40 infilling angles in different regions to encode data.
  }
  \label{fig:highcapacitytag}
  \end{center}
\end{figure}

As mentioned in Section~\ref{AnisoTag_setting}, dividing AnisoTag into several parallel regions and applying different 3D printing settings based on our \texttt{G-code} tool provides AnisoTag with a high information capacity. Moreover, the capacity would be further improved if AnisoTag is extracted with image processing operations similar to existing 3D printed tagging methods. Specifically, we could extend the encoding region division to two dimensions and obtain more encoding blocks; see Figure~\ref{fig:highcapacitytag}. Compared with methods whose information capacity is limited by the parameters available in the 3D model slicing software, e.g., G-ID~\cite{G-id}, AnisoTag increases the information capacity of 3D printed tags by a factor of tens.

\section{Conclusion, Limitation, and Future Work}
We have presented AnisoTag, a lightweight and low-cost machine-readable tagging system on 3D printed 2D surfaces, which could be fabricated using a consumer-grade FDM 3D printer and various types and colors of 3D printing materials. According to supplement experiments in Appendix, AnisoTag also has the robustness to daily usage abrases and works well under a variety of light conditions. The principle and techniques of AnisoTag could be adopted in other applications and have produced some results worth discussing (Section~\ref{sec:extend}).

AnisoTag is currently available on 3D printed surfaces and objects with simple shapes. Benefiting from lower requirements for the precision of extraction device and complexity of decoding algorithm, embedding data on the planar surface of 3D printed objects is not unusual (Table~1), especially when some categories of 3D models, e.g., figurines and CAD \cite{layer_thickness}, commonly have planar surfaces. The performance of AnisoTag is mainly limited by the capability of the hardware used in the detection prototype. As discussed in our experiments, the extraction accuracy depends on the sensitivity and number of optical sensors set on the background plane. The information capacity is affected by the diameter of the laser source and the sensitivity of the optical sensor. The detection prototype with a finer laser source and more sensitive optical sensors could detect AnisoTag with more encoding regions and encoded angles, i.e., AnisoTag with a larger information capacity. And the detection speed relies on the decoding algorithm and the speed of the microcontroller converting and transmitting the data obtained from sensors. Therefore, for higher demanding application scenarios, AnisoTag could still work well with the corresponding hardware-upgraded detection prototype.

Moreover, we notice that the current design of the detection prototype is not stable enough in some special situations. For example, in the preliminary decoding method, the region discrimination is based on the weaker reflection of the narrow borderline between two encoding regions, which limits the swiping speed, i.e., the detection efficiency. Then, when the intensity of illumination changes unevenly, the resistance value of photoresistors would change inconsistently, causing interferences with the detection performance. In addition, the detection prototype requires the relative positions among the laser source, swiping region of AnisoTag, and the background plane to remain constant. As a result, the current prototype design cannot work in unstable situations, e.g., on running trains or hands. Last but not least, the laser pointer pen is not suitable as a laser source for long-term use, whose illumination intensity varies with its battery power. So we plan to optimize the detection strategy and the hardware design of the prototype in future work. Specifically, we could use a laser source connected to the circuit to maintain a constant laser intensity and fix the laser source and the background plane with relative positions in a compact dark box. The dark box would have one opening or notch for swiping AnisoTag. This design is expected to address the above shortcomings.

We also wonder how the other light sources reflect on the SCS microstructure, with the most noteworthy one being the line laser source. AnisoTag irradiated by line laser would reflect all the illumination patterns simultaneously, which contributes to designing a faster AnisoTag detection method; see reflection image in the supplementary material. So another interesting future direction is how to design a detection process for AnisoTag under the line laser. 

%% file: sections/appendix.tex
\section{The Proof of Reflected Light Cone}
The proof process still follows the coordinate system established in Figure~\ref{coordinate_system}(b). By associating Eq.~(\ref{reflected_vector}), the reflected ray direction $\vec{b}$ can be re-expressed as:
\begin{equation}
\begin{aligned}
    \Gamma &= \vec{n} \cdot \vec{a} = \sin\alpha \sin\theta \sin\phi-\cos\alpha \cos\theta\\
    \vec{b} &= \vec{a} - 2\Gamma\vec{n} \\
    &= (-2\Gamma\sin\theta \cos\phi, \sin\alpha-2\Gamma\sin\theta\sin\phi, -\cos\alpha-2\Gamma\cos\theta).
\end{aligned}
\end{equation}
According to the law of reflection, $|\vec{b}|=|\vec{a}|=1$. As mentioned in Section~\ref{anisotropy_analysis_2}, the cylindrical axis direction is $\vec{c} = (\cos\delta, \sin\delta,0) = (-\sin\phi, \cos\phi, 0)$. Defining the angle between $\vec{b}$ and $\vec{c}$ as $\xi$, $\cos\xi$ could be calculated as:
\begin{equation}
    \begin{aligned}
    \cos\xi &= \frac{\vec{b}\cdot\vec{c}}{|\vec{b}||\vec{c}|} = \vec{b}\cdot\vec{c}\\
    &= 2\Gamma\sin\theta\sin\phi\cos\phi + \cos\phi\sin\alpha - 2\Gamma\sin\theta\sin\phi\cos\phi\\
    &= \cos\phi\sin\alpha.
    \end{aligned}
\end{equation}
As a result, for a given incident angle $\alpha$ and cylindrical section angle $\phi$, the reflected ray direction $\vec{b}$ changes with $\theta$ but maintains a fixed angle with the cylindrical axis. Thus, the reflected rays form a circular cone.

\begin{figure}[t]
\setlength{\belowcaptionskip}{0pt}
\setlength{\abovecaptionskip}{0pt}
  \begin{center}
  \includegraphics[width=3.1in]{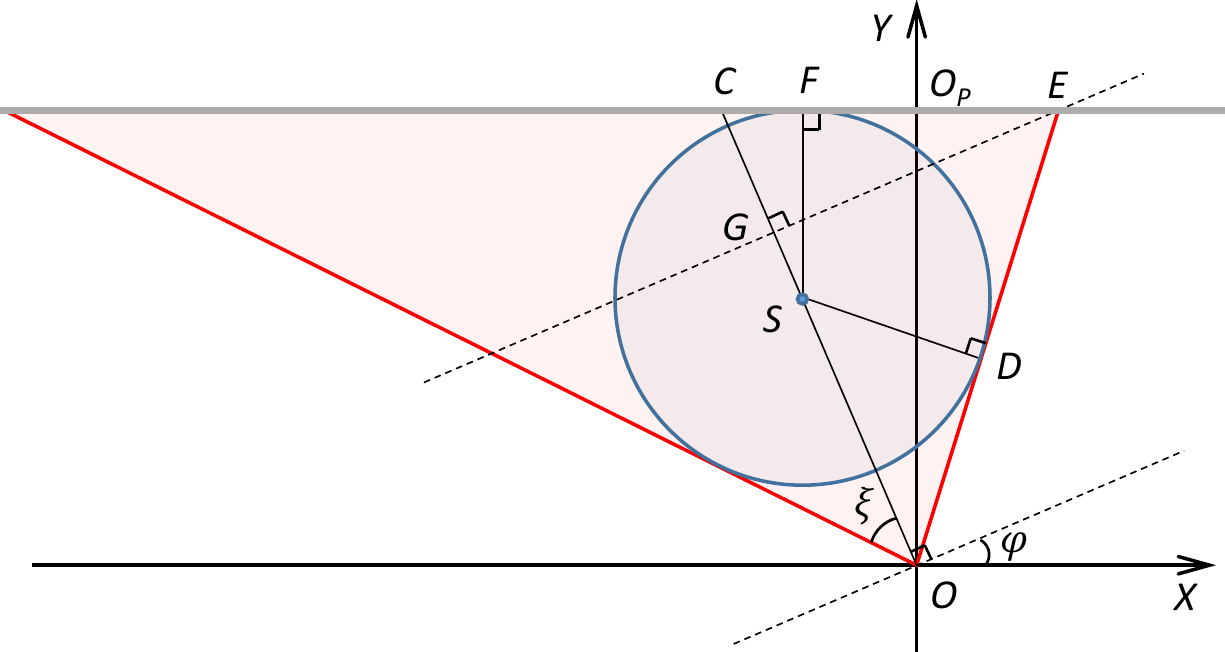}
  \caption{A top view of a reflection model with a Dandelin sphere tangent internally to the light cone and the background plane.
  }\label{dandelin_sphere}
  \end{center}
\end{figure}

\begin{figure}[t]
\setlength{\belowcaptionskip}{0pt}
\setlength{\abovecaptionskip}{3pt}
  \begin{center}
  \includegraphics[width=3in]{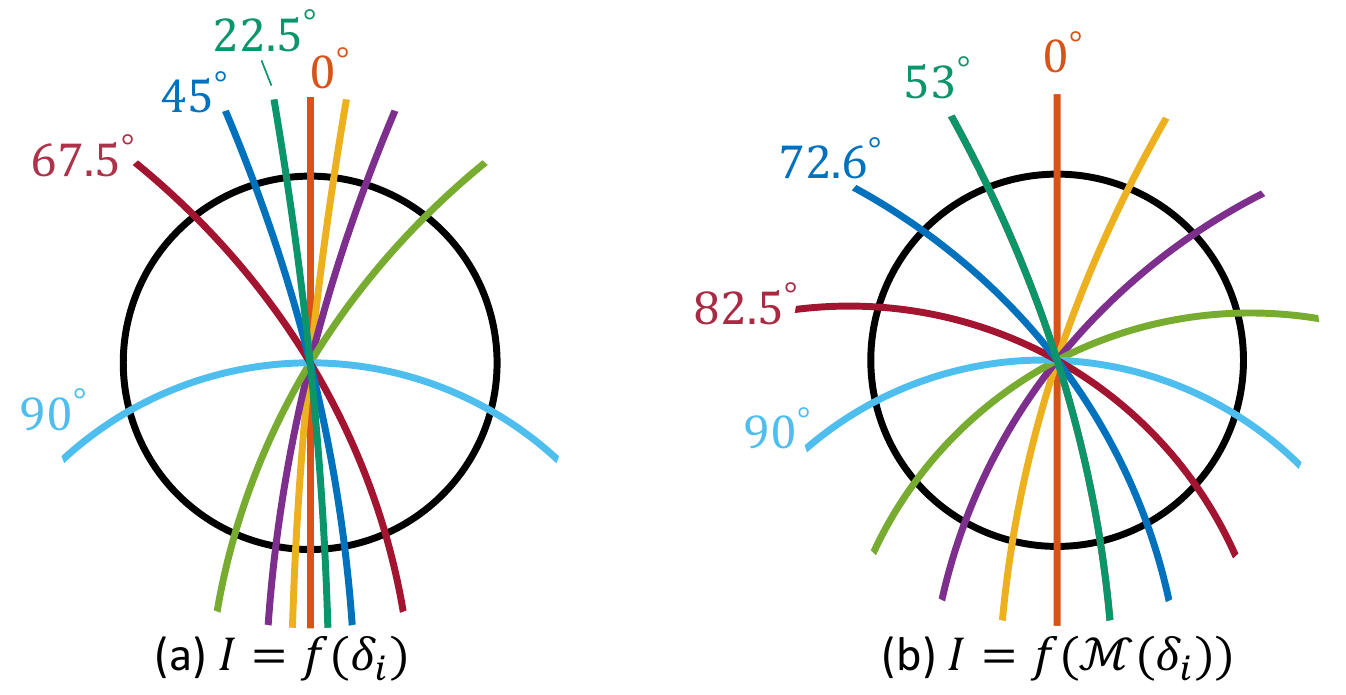}
  \caption{\textbf{Illumination patterns $\mathcal{I}$ reflected by SCS microstructures of different angles.} These patterns pass through the same fixed point. And optical sensors would be placed on the circumference of the black circle. In this example, the SCS microstructures encode $3$ bits, having $8$ encoding angles. In (a), the encoding angles $\delta_i$ are uniformly distributed, but they reflect unevenly distributed illumination patterns. And in (b), the encoding angles are remapped by the nonlinear mapping function $\mathcal{M}(\cdot)$, resulting in reflected illumination patterns with similar intervals.
  }\label{nolinear_mapping}
  \end{center}
\end{figure}

\section{Function of Reflected illumination Pattern}
The reflected light cone intersects with the background plane and forms the reflected illumination patterns $\mathcal{I}$, which is a conic section. To determine the function parameters of the reflected illumination pattern, we use one inference about conic sections named Dandelin sphere \cite{Dandelin_book,Dandelin_wiki}. Specifically, inside the reflected lighting cone, we draw a sphere tangent to that cone and the background plane $P: y=d$. Figure~\ref{dandelin_sphere} shows the plane view of the reflected lighting cone and the sphere observed along the direction of the inverse $Z$-axis. According to the inference of the Dandelin sphere, the eccentricity $e$ of the reflected illumination pattern is:
\begin{equation}
    e = \frac{\sin \angle CEG}{\sin \angle OEG} = \frac{\sin \phi}{\cos \xi},
\end{equation}
where $\phi$ is the cylindrical section angle and $\xi$ is the half cone angle calculated in the previous section. The sphere is tangent to the background plane at point $F$, which is the focus of the reflected illumination pattern. To obtain the coordinates of $F$, we first calculate the length of $FO_p$:
\begin{equation}
\label{eq1}
    FO_p=\frac{FO_p}{CO_p}CO_p=\frac{SO}{CO}CO_p=\frac{SO}{CS+SO}CO_p.
\end{equation}
The ratio relation between $CS$ and $SO$ could be derived by:
\begin{equation}
\label{eq2}
    CS\cos\phi=SF=SD=SO\sin\xi.
\end{equation}
Substituting Eq.~(\ref{eq2}) to Eq.~(\ref{eq1}), the length of $FO_p$ could be calculated by:
\begin{equation}
\begin{aligned}
    FO_p&=\frac{\cos\phi}{\cos\phi+\sin\xi}CO_p\\
    &=\frac{\cos\phi}{\cos\phi+\sin\xi}\tan\phi \cdot OO_p\\
    &=\frac{\cos\phi}{\cos\phi+\sin\xi}\tan\phi \cdot d.
\end{aligned}
\end{equation}
Then, we obtain the coordinates of $F$:
\begin{equation}
F = \left(-\frac{\tan \phi \cos\phi}{\cos\phi+\sin\xi}d,\ d,\ 0 \right).
\end{equation}
With the eccentricity and the focus, the function of the reflected illumination pattern $\mathcal{I}=f(\alpha,\phi,d)$ is easy to determine.

\section{Nonlinear Mapping Function}
As mentioned in Section~\ref{encoding}, there is a nonlinear mapping function $\mathcal{M}(\cdot)$ remapping the angular intervals to reflect more easily detectable illumination patterns with similar intervals. Taking Figure~\ref{nolinear_mapping}(a) as an example, it is more difficult to distinguish the illumination patterns reflected from SCS microstructures with cylindrical axis angles $0^\circ$ and $22.5^\circ$ than those with angles $67.5^\circ$ and $90^\circ$. The reason is that the mapping between SCS microstructure angles and these intersections is nonlinear. That nonlinear mapping function could be calculated based on the reflected illumination function $\mathcal{I}=f(\alpha, \phi, d)$ and the experimental settings, and its inverse function is set as the remapping function $\mathcal{M}(\cdot)$ mentioned in Eq.~(\ref{angle_mapping}). As a result, the nonlinear part of the reflection process would be eliminated and the SCS microstructures could reflect illumination patterns with uniform intervals; see Figure~\ref{nolinear_mapping}(b). In summary, we can deploy optical sensors along a ``detection circle'' around the fixed point to detect the intersection points of the reflected illumination patterns with that circle, and thus determine the state of the illumination pattern. The data on AnisoTag are then extracted based on the mapping relationship.

\begin{figure}[t]
\setlength{\belowcaptionskip}{0pt}
\setlength{\abovecaptionskip}{1pt}
    \begin{center}
    \includegraphics[width=3.3in]{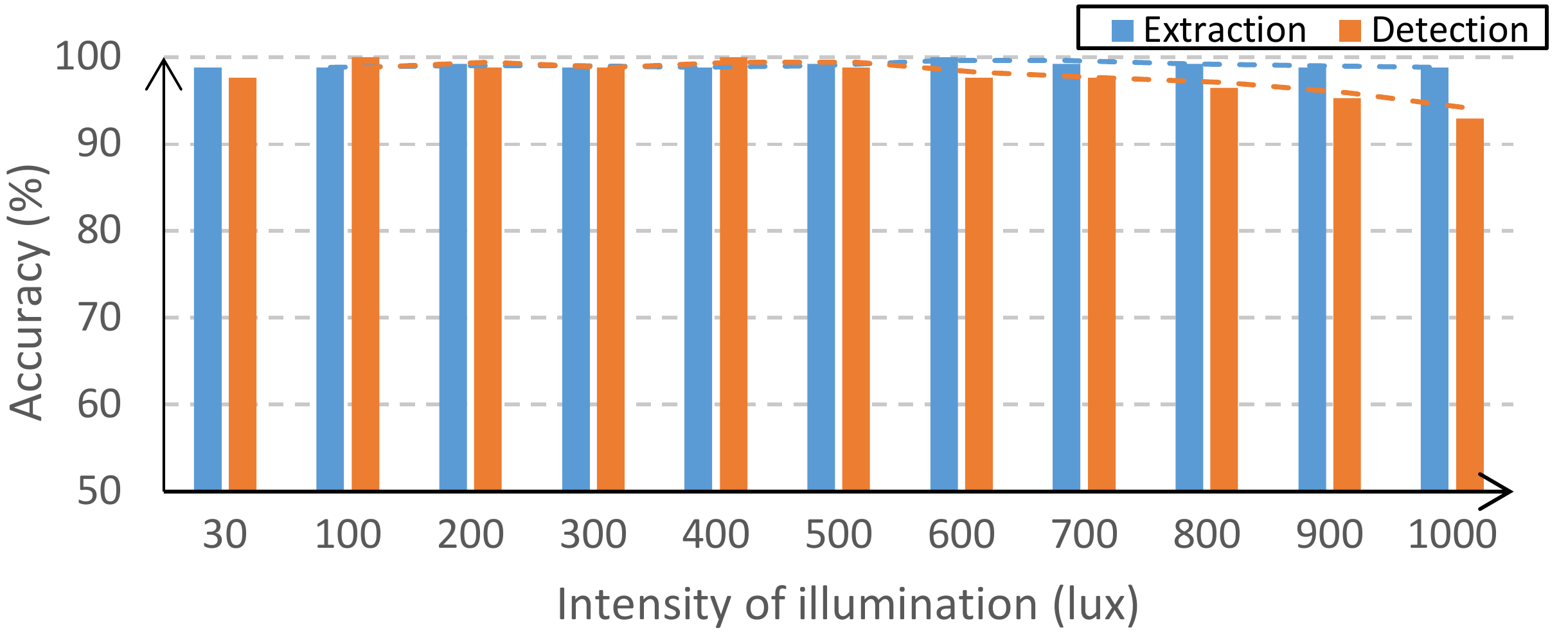}
    \caption{The impact of light conditions on the data detection and extraction accuracy of AnisoTag.
    }\label{fig:lux}
    \end{center}
\end{figure}

\section{Supplementary Experiments}
\subsection{Light Conditions}
To understand how light conditions affect the detection results of AnisoTag, we use a dimmable LED light source to adjust the ambient illumination intensity around the detection prototype from $30$ to $1000$ lux. First, we perform the pre-processing operation of the detection process, i.e., recording the voltage signals of the detection prototype under the default ambient illumination intensity $400 lux$. Then, under each light condition, we perform the detection process $5$ times and calculate the average accuracy. According to the trendline (dashed lines) shown in Figure~\ref{fig:lux}, the variation of ambient illumination intensity has few impacts on data extraction accuracy, yet the data detection accuracy decreases when the intensity of illumination exceeds $700 lux$, which could be explained as follows. The variation in the illumination intensity changes the voltage value of the photoresistors on the detection prototype, which can be roughly considered as adding or subtracting a constant from the original one. After the normalization operation in the detection process, the variation of illumination intensity affects a little the determination of the reflected illumination pattern. However, as the illumination intensity increases, the change in the resistance of the photoresistor slows down, resulting in less voltage change when it is illuminated by reflected illumination patterns. So there is interference with the detection results when the illumination intensity is above a certain value. However, under the illumination intensity of the common working occasions ($300$\textasciitilde$800 lux$) and the non-working occasions ($50$\textasciitilde$100 lux$), AnisoTag performs well; see Figure~\ref{fig:lux}.

\subsection{Durability}
In this paper, to facilitate experiments, we use a single 3D printed layer of $0.2mm$ thickness to fabricate each AnisoTag, which makes it bendable in some video and figure presentations. But in real application scenarios, the AnisoTag would be combined with 3D printed object, thus the AnisoTag and the product should be considered as a whole to evaluate structural durability. Moreover, the slightly above-surface borderline on AnisoTag mentioned in Section~\ref{sec:detection_prototype} could protect the SCS microstructure from daily usage abrases. Experimentally, moving the AnisoTag by $1m$ on $1000$ grit sandpaper with $10$ times its self-weight, its reflected illumination pattern has the same shape and less than $5\%$ illumination intensity change. Additionally, the material universality of AnisoTag in the experiments allows us to fabricate it with materials like Nylon for scenarios with a need for higher durability.